\documentclass[12pt] {article}
\pdfoutput=1
\usepackage[hyphens,spaces,obeyspaces]{url}
\usepackage{setspace}

\usepackage{epsfig}
\usepackage{amssymb}
\usepackage{amsmath}
\usepackage{multirow}
\usepackage{setspace}
\usepackage{datetime}
\usepackage{amsmath}
\usepackage{algorithm}
\usepackage{algpseudocode}
\usepackage{graphicx}
\usepackage{subfig} % for subfigures
\usepackage{lipsum}
\usepackage{acronym}
\usepackage{array}
\usepackage{graphicx}
\usepackage{cite}
\usepackage{graphicx}
\usepackage{epstopdf}
\usepackage{caption}
\usepackage{balance}
\usepackage{fancybox}
\usepackage{colortbl}
\usepackage{array}
\usepackage{mystyle}
\usepackage{multirow}
\usepackage[obeyspaces,hyphens,spaces]{url}
\usepackage{threeparttable}
\usepackage[english]{babel}% tables
\usepackage{longtable}
\usepackage{listings, lstautogobble}
\usepackage{booktabs}
\usepackage{threeparttable}
\usepackage[table,xcdraw]{xcolor}
\usepackage{amsthm}
\theoremstyle{definition}
\newtheorem{thm}{Theorem}%[section] % reset theorem numbering for each section
\newtheorem{defn}{Definition}%[section] % definition numbers are dependent on theorem numbers
\AtBeginEnvironment{thm}{\footnotesize}
\AtBeginEnvironment{defn}{\footnotesize}
\renewcommand{\baselinestretch}{1}

\providecommand{\keywords}[1]{\textbf{\textit{Keywords---}} #1}

{\begin{list}{}%
         {\setlength{\leftmargin}{#1}}%
         \item[]%
}
{\end{list}}

\newdateformat{monthyeardate}{%
  \monthname[\THEMONTH], \THEYEAR}
\begin{document}
\title{\Large{\textbf{Dynamic Fault Trees Analysis using an Integration of Theorem Proving and Model Checking} }}
\author{
Yassmeen~Elderhalli$^{1}$, Osman~Hasan$^{1,2}$, Waqar~Ahmad$^{2}$ \\ and Sofi\`ene~Tahar$^{1}$ \vspace*{2em}\\
$^{1}$Department of Electrical and Computer Engineering,\\
Concordia University, Montr\'eal, Canada  \\
\{y\_elderh,o\_hasan,tahar\}@ece.concordia.ca \vspace*{2em}\\
$^{2}$Electrical Engineering and Computer Science,\\
National University of Science and Technology, Islamabad, Pakistan \\
waqar.ahmad@seecs.nust.edu.pk \vspace*{3em}\\
\textbf{TECHNICAL REPORT}\\
\date{December 2017}
}
\maketitle

\newpage
\begin{abstract}
Dynamic fault trees (DFTs) have emerged as an important tool for capturing the dynamic behavior of system failure. These DFTs are then analyzed qualitatively and quantitatively using stochastic or algebraic methods to judge the failure characteristics of the given system in terms of the failures of its sub-components. Model checking has been recently proposed to conduct the failure analysis of systems using DFTs with the motivation to provide a rigorous failure analysis of safety-critical systems. However, model checking has not been used for the DFT qualitative analysis and the reduction algorithms used in model checking are usually not formally verified. Moreover, the analysis time grows exponentially with the increase of the number of states. These issues limit the usefulness of model checking for analyzing complex systems used in safety-critical domains, where the accuracy and completeness of analysis matters the most. To overcome these limitations, we propose a comprehensive methodology to perform the qualitative and quantitative analysis of DFTs using an integration of theorem proving and model checking based approaches. For this purpose, we formalized all the basic dynamic fault tree gates using higher-order logic based on the algebraic approach and formally verified some of the simplification properties. This formalization allows us to formally verify the equivalence between the original and reduced DFTs using a theorem prover, and conduct the qualitative analysis. We then use model checking to perform the quantitative analysis of the formally verified reduced DFT. We applied our methodology to five benchmarks and the results show that the formally verified reduced DFT was analyzed using model checking with up to six times less states and up to 133000 times faster.

\end{abstract}

\newpage
\tableofcontents
\newpage
\section{Introduction}
\label{introduction}
A fault tree (FT) \cite{DFT-survey} is a graphical representation of the causes of failure of a system that is usually represented as the top event of the fault tree. FTs can be categorized as Static Fault trees (SFT) and Dynamic Fault trees (DFT) \cite{DFT-survey}. In SFT, the structure function (expression) of the top event describes the failure relationship between the basic events of the tree using FT gates, like AND and OR, without considering the sequence of failure of these events. DFTs, on the other hand, model the failure behavior of the system using dynamic FT gates, like the spare gate, which can capture the dependent behavior of the basic events along with the static gates. DFTs provide a more realistic representation of systems using the dynamic gates. For example, the spare DFT gate can model the failure of the car tires and their spares that cannot be modeled using the SFT gates. \\ 
\indent Fault Tree Analysis (FTA) \cite{DFT-survey} has become an essential part of the safety-critical system design process, where the causes of failure and their probabilities should be considered at an early stage. There are two main phases for FTA, the qualitative analysis and the quantitative analysis \cite{DFT-handbook}. In the \textit{qualitative analysis}, the cut sets and cut sequences are determined, which, respectively, represent combinations and sequences of basic events of the DFT that cause a system failure \cite{DFT-survey}. The \textit{quantitative analysis} provides numeric analysis results about the probability of failure of the top event and the mean time to failure among other metrics \cite{DFT-survey}. Dynamic FTA is commonly done  algebraically \cite{Merle-thesis} and using Markov chains \cite{DFT-handbook}. In the algebraic approach, an algebra similar to the Boolean algebra is used to determine the structure function of the top event. Based on this algebra, the structure function can be reduced to determine a reduced form of the cut sets and sequences. The probabilistic analysis of the FT can then be performed based on the reduced form of the generated structure function by considering the probability of failure of the basic events. For the Markov chain based analysis, the FT is first converted to its equivalent Markov chain and then the probability of failure of the top event is determined by analyzing the generated Markov chain. The resultant Markov chain can be very large, while dealing with complex systems, which limits the usage of Markov chains in DFT analysis.\\
%
%The FT is converted to a Markov chain by starting with an initial working state, then the chain is created by failing each input of the tree until all the possible combination and sequences are recorded in the chain\cite{DFT-handbook}. . A modular approach can be followed, which divides the tree into two parts, static and dynamic. The static part can be analyzed using the traditional static fault tree analysis methods, while the dynamic tree can be analyzed using Markov chains\cite{Modular-DFT}. 
\indent Traditionally, the dynamic FTA is performed using paper-and-pencil proof methods or computer simulation. While the former is error prone, specially for large systems, the latter provides a more scalable alternative. However, the results of simulation cannot be termed as accurate due to the involvement of several approximations in the underlying computation algorithms and the sampling based nature of this method. Given the dire need of accuracy in failure analysis of safety-critical systems, formal methods \cite{DFT-survey} have also been recently explored for DFT analysis. For example, the STORM probabilistic model checker \cite{dehnert2017storm} has been used to analyze DFTs based on Markov chain analysis \cite{volk2017fast}. Similarly, higher-order logic (HOL) theorem proving has been used to formalize and analyze SFTs \cite{ahmed2016formalization}. However, probabilistic model checking has not been used in the formal qualitative analysis of DFTs. Moreover, it cannot support the analysis of large systems unless a reduction algorithm is invoked, and the implementation of such reduction is usually not formally verified. This means that one cannot ascertain that the analysis results after reduction are accurate or correspond to the original system. On the other hand, the only support for FTs in HOL is limited to SFTs.\\
% by converting this DFT into its equivalent Markov chain. Using probabilistic model checkers in DFT analysis is promising, since the analysis is automated. However, generally large systems cannot be analyzed due to the large state space that is generated from these systems. Theorem proving, such as HOL4\cite{HOL4}, can be used to analyse DFTs, which represents an expressive and scalable method that can handle large systems. However, theorem provers are interactive, which means that they require the guidance of the verification engineer and cannot be fully automated. 
\indent We propose to overcome the above-mentioned limitations of formal DFT analysis by using an integrated model checking and theorem proving based methodology. We propose to use theorem proving for verifying the equivalence between the original and the reduced form of the DFT. The formally verified reduced DFT can then by quantitatively analyzed using model checking. Thus, the proposed methodology tends to provide a more sound analysis than the sole model checking based analysis due to the involvement of a theorem prover in the verification of the reduced model. Moreover, it caters for the state-space based issues of model checking by providing it a reduced model for the quantitative analysis. The foremost components of the proposed methodology include the formalization of the dynamic gates and their formally verified reduction theorems, which in turn are used to verify the equivalence between the original DFTs and the reduced ones. Using this verified reduced DFT, a reduced form of the cut sets and sequences of the structure function of the DFT can be formally verified within a theorem prover. We then perform the quantitative analysis of the formally verified reduced DFT in model checking and thus reduce the generated state space and the analysis time. More importantly, we are confident that the analysis results of the reduced DFT correspond to the original DFT, as the reduction is verified using theorem proving. In order to illustrate the utilization and effectiveness of the proposed methodology, we analyzed five benchmark DFTs, i.e., a Hypothetical Example Computer System (HECS) \cite{DFT-handbook}, a Hypothetical Cardiac Assist System (HCAS) \cite{boudali2007compositional , Merle-thesis}, a scaled cascaded PAND DFT \cite{boudali2007compositional , MerlePAND}, a multiprocessor computing system \cite{TrivediMCS1995 , boudali2007compositional} and a variant of the Active Heat Rejection System (AHRS) \cite{Bayesian-Dugan}.

The reduced DFTs and their reduced cut sequences are formally verified using HOL4 theorem prover. In addition, each DFT is analyzed twice using STORM model checker, one without any reduction and the other using the reduced DFTs. The analysis results show that using the verified reduced DFT for the quantitative analysis allows us to reduce the number of generated states by the model checker and the time required to perform the analysis.

The rest of the report is structured as follows: Section \ref{relatedwork} presents some related work. Section \ref{methodology} provides a detailed description of the proposed methodology. In Section \ref{Formalization_in_hol}, we present our HOL formalization of DFT gates. In Section \ref{Simplification}, we provide the details of the verification of the simplification theorems. Section \ref{Experiment results} describes a set of experimental results. Finally, we conclude the report in Section \ref{Conclusion}.

\section{Related Work}
\label{relatedwork}
DFT analysis has been conducted using various tools and techniques\cite{DFT-survey}. For example,  Markov chains have been extensively used for the modeling and analysis of DFTs  \cite{DFT-handbook}. The scalability of Markov chains in analyzing large DFTs is achieved by using a modularization approach \cite{Modular-DFT}, where the DFT is divided into two parts: static and dynamic. The static subtree is analyzed using the ordinary SFT analysis methods, such as Binary Decision Diagrams (BDD) \cite{DFT-survey}, and the dynamic subtree is analyzed using Markov chains. This kind of modularization approach is available in the Galileo tool \cite{Galileo}. In \cite{boudali2007compositional}, the authors use a compositional aggregation technique to develop Input-Output Interactive Markov Chains (I/O-IMC) to analyse DFTs. This approach is implemented in the DFTCalc tool \cite{arnold2013dftcalc}. The algebraic approach has also been extensively used in the analysis of DFTs \cite{Merle-thesis}, where the top event of the DFT can be expressed and reduced in a manner similar to the ordinary Boolean algebra. The reliability of the system expressed algebraically can be evaluated based on the algebraic expression of the top event \cite{MerlePAND}. The main problem with the Markov chain analysis is the large generated state space when analyzing complex systems, which requires high resources in terms of memory and time. Moreover, simulation is usually utilized in the analysis process, which does not provide accurate results. Although modularization tends to overcome the large state-space problem with Markov chains,  we cannot obtain a verified reduced form of the cut sequences of the DFT. The algebraic approach provides an algebraic framework for performing both the reduction and the analysis of the DFT. However, the foundations of this approach have not been formalized, which implies that the results of the analysis should not be relied upon especially in safety-critical systems.\\
\indent Formal methods can overcome the above-mentioned inaccuracy limitations of traditional DFT analysis techniques. Probabilistic model checkers, such as STORM\cite{dehnert2017storm}, have been used for the analysis of DFTs. The main idea behind this approach is to automatically convert the DFT of a given system into its corresponding Markovian model and then analyze the safety characteristics quantitatively of the given system using the model checker \cite{ghadhab2017model}. The STORM model checker accepts the DFT to be analyzed in the Galileo format \cite{Galileo} and generates a failure automata of the tree. This approach allows us to verify failure properties, like probability of failure, in an automatic manner. However, the approach suffers from scalability issues due to the inherent state-space explosion problem of model checking. Moreover, the implementation of the reduction algorithms used in model checkers are generally not formally verified. Finally, model checkers have only been used in the context of probabilistic analysis of DFTs and not for the qualitative analysis, as the cut sequences in the qualitative analysis cannot be provided unless the state machine is traversed to the fail state, which is difficult to achieve  for large state machines. \\
\indent Exploiting  the expressiveness of higher-order logic (HOL) and the soundness of theorem proving, Ahmad et.al \cite{ahmed2015towards , ahmed2016formalization} formalized static fault trees in HOL4 and evaluated the probability of failure based on the Probabilistic Inclusion-Exclusion principle. However, the main problem in theorem proving lies in the fact that it is interactive, i.e., it needs user guidance in the proof process. Moreover, to the best of our knowledge, no higher-order-logic formalization of DFTs is available in the literature so far and thus it is not a straightforward task to conduct the DFT analysis using a theorem prover as of now.\\
\indent It can be noted that both model checking and HOL theorem proving exhibit complementary characteristics, i.e., model checking is automatic but cannot deal with large systems and does not provide qualitative analysis of DFTs, while HOL theorem proving allows us to verify universally quantified generic mathematical expressions but at the cost of user interventions. In this work, we leverage upon the complementary nature of these approaches to present an integrated methodology that provides the expressiveness of higher-order logic and the existing support for automated probabilistic analysis of DFTs using model checking. The main idea is to use theorem proving to formally verify the equivalence between the original and the reduced DFT and then use a probabilistic model checker to conduct quantitative analysis on the reduced DFT. As a result, both the generated state machine and the analysis time are reduced. In addition, a formally verified reduced form of the cut sequences is also obtained.
\section{Proposed Methodology}
\label{methodology}

\begin{figure}[]
\centering
%\vspace{-16pt}
\includegraphics[scale=0.5]{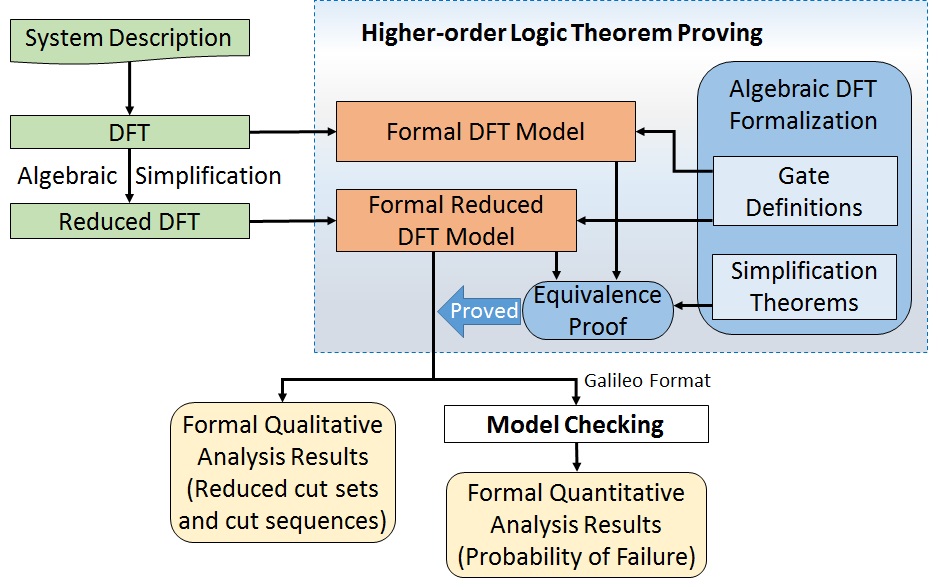}
%\vspace{-8pt}
\caption{Overview of Proposed Methodology}
%\vspace{-16pt}
\label{fig:methodology_fig}
\end{figure}

The proposed methodology for the formal DFT analysis is depicted in Figure \ref{fig:methodology_fig}. It provides both formal DFT qualitative analysis using theorem proving and quantitative analysis using model checking. The DFT analysis starts by having a system description. The failure behavior of this system is then modeled as a DFT, which can be reduced based on the algebraic approach \cite{Merle-thesis}. The idea of this algebraic approach is to deal with the events, which can represent the basic events or outputs, according to their time of failure ($d$). For example, $d$($X$) represents the time of failure of an event $X$. In the algebraic approach, temporal operators (Simultaneous (\(\Delta\)), Before (\(\lhd\)) and Inclusive Before (\(\unlhd\))) are defined to model the dynamic gates. Based on these temporal operators, several simplification theorems exist to perform the required reduction. This reduction process can be erroneous if it is performed manually using paper-and-pencil. Moreover, reduction algorithms may also provide wrong results if they are not formally verified. In order to formally check the equivalence between the original model and the reduced one, we developed a library of formalized dynamic gates in HOL and verified their corresponding simplification theorems. These foundations allow us to develop a formal model for any DFT using the formal gate definitions. Based on the verified simplification theorems, we can then verify the equivalence between the formally specified original and the reduced DFT models using a theorem prover. The formally verified reduced structure function can then be utilized to perform the qualitative analysis of the reduced model in the theorem prover as well as its quantitative analysis by using a model checker.

The qualitative analysis represents an important and a crucial step in DFT analysis, since it allows to identify the sources of failure of the system without the availability of any information or actual numbers about the failure probabilities of the basic events. In static fault trees, the qualitative analysis is performed by finding the cut sets. Due to the temporal behavior of the dynamic gates, just finding the cut sets does not capture the sequence of failure of events that can cause the system failure. The cut sequences on the other hand capture not only the combination of basic events but also the sequence of events that can cause the system failure. In the proposed methodology, a theorem prover is used to verify a reduced expression of the structure function of the top event, which ensures that the reduction process is accurate. Using this reduced structure function, a formally verified reduced form of the cut sequences can also be determined. 

The formally verified cut sets and sequences for the DFT and a reduced form of the structure function of the top event can now be used in a probabilistic model checker to do the quantitative analysis of the given system. Because of the reduced model, we get a reduction in the analysis time and number of states. In this work, the STORM model checker is used to perform the probabilistic analysis of the DFT. Several input languages are supported by this model checker including the Galileo format for DFT. Both the probability of failure of the top event as well as the mean time to failure can be computed using STORM. It is worth mentioning that since the analyzed model of the DFT is a Markov Automata (MA) (in case of non-deterministic behavior) or a Continuous Time Markov Chain (CTMC), only exponential failure distributions are supported by the proposed methodology.  
\section{Formalization of Dynamic Fault Trees in HOL}
\label{Formalization_in_hol}
In this section, we present the formal definitions in HOL of the identity elements, the temporal operators and the dynamic gates. It is assumed that a fault is represented using an event. The occurrence of a fault indicates that the corresponding event is true. It is also assumed that the events are non-repairable. 
%\vspace{-20pt}
\subsection{Identity Elements}

Two identity elements are defined, these are the \emph{ALWAYS} and the \emph{NEVER} elements. The \emph{ALWAYS} identity element represents an event with a time of failure equals to 0. The \emph{NEVER} element represents an event that never occurs. These two elements are defined based on their time of failure in HOL as follows:\\ 
\begin{defn}
%\texttt{\bf{Definition 3: }}
\label{ALWAYS_def}
\vspace{-16pt}
\emph{ALWAYS element} \\
\vspace{1pt} {\texttt{$\vdash$ ALWAYS = (0:extreal)}}\\

\end{defn}
\begin{defn}
%\texttt{\bf{Definition 3: }}
\label{NEVER_def}
\vspace{-16pt}
\emph{NEVER element} \\
\vspace{1pt} {\texttt{$\vdash$ NEVER = PosInf}}\\
\end{defn}
%
%\noindent\label{ALWAYS_NEVER} \vspace{-8pt}
%\begin{alltt}
%\(\vdash\) ALWAYS = (0:extreal) 
%\(\vdash\) NEVER = PosInf 
%\end{alltt}
%\noindent 
\vspace{-16pt}
\noindent where \texttt{extreal} is the HOL data type for extended real numbers, which includes positive infinity (+\(\infty\)) and negative infinity (-\(\infty\)) and \texttt{PosInf} is the (+\(\infty\)) representation in HOL.
%\vspace{-8pt}
\subsection{Temporal Operators}
We formalize three temporal operators to model the dynamic behavior of the DFT: \emph{Simultaneous} (\(\Delta\)), \emph{Before} (\(\lhd\)) and \emph{Inclusive Before} (\(\unlhd\)). The \emph{Simultaneous} operator has two input events, which represent basic events or subtrees. The time of occurrence (failure) of the output event of this operator is equal to the time of occurrence of the first or the second input event considering that both input events occur at the same time:
\begin{equation}
d(A  \Delta B) =  \begin{cases} d(A), &d(A) = d(B)\\ +\infty,  &d(A)  \neq d(B) 
\end{cases} 
\end{equation}
It is assumed that for any two basic events, if the failure distribution of the random variables that represent these basic events is continuous then they cannot have the same time of failure, and hence the result of the \emph{Simultaneous} operator between them is \emph{NEVER}.
\vspace{-10pt} 
\begin{equation}
d(A  \Delta B) =  NEVER
\end{equation}
\noindent where $A$ and $B$ are basic events with random variables that exhibit continuous failure distributions.
 
The \emph{Before} operator accepts two input events, which can be basic events or two subtrees. The time of occurrence of the output event of this operator is equal to the time of occurrence of the first input event if the first input event (left) occurs before the second input event (right), otherwise the output never fails:
\begin{equation}
d(A  \lhd B) =  \begin{cases}d(A), &d(A) < d(B)\\ +\infty, &d(A)\geq d(B) 
\end{cases} 
\end{equation}
The \emph{Inclusive Before} combines the behavior of both the \emph{Simultaneous} and \emph{Before} operators, i.e., if the first input event (left) occurs before or at the same time as the second input event (right), then the output event occurs with a time of occurrence equal to the time of occurrence of the first input event:
\begin{equation}
d(A  \unlhd B) =  \begin{cases}d(A), &d(A) \leq d(B)\\ +\infty, &d(A) > d(B)
\end{cases} 
\end{equation}

\noindent We formalize these temporal operators in HOL as follows:
\begin{defn}
\label{SIMULT_def}
\emph{Simultaneous Operator} \\
\vspace{1pt} {\texttt{$\vdash$ $\forall$ (A:extreal) B.
		D\_SIMULT A B = if (A = B) then A else PosInf
}}
\end{defn}
%\vspace{-8pt}
\begin{defn}
\label{Before_def}
\emph{Before Operator} \\
\vspace{1pt} {\texttt{$\vdash$ $\forall$ (A:extreal) B. 
		D\_BEFORE A B = if (A < B) then A else PosInf
}}
\end{defn}
%\vspace{-8pt}
\begin{defn}
\label{Inclusive_Before_def}
\emph{Inclusive Before Operator} \\
\vspace{1pt}\noindent{\texttt{$\vdash$ $\forall$ (A:extreal) B. D\_INCLUSIVE\_BEFORE A B = if (A \(\leq\) B) then A                 else PosInf
}}
\end{defn}
%\vspace{-4pt}
\noindent where $A$ and $B$ represent the time of failure of the events $A$ and $B$, respectively. 
%\vspace{-20pt}
\subsection{Fault Tree Gates}
\begin{figure}[]
\centering
%\vspace{-40pt}
\subfloat[AND]{\label{fig:AND}{\includegraphics[width=0.12\textwidth , height=3cm]{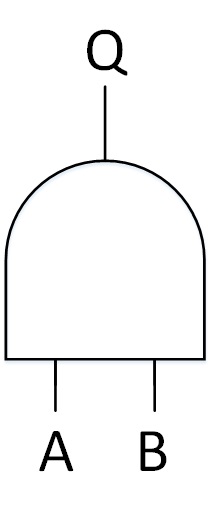}}}\hfill \hfill \hfill \hfill \hfill 
\subfloat[OR]{\label{fig:OR}{\includegraphics[width=0.12\textwidth , height=3cm]{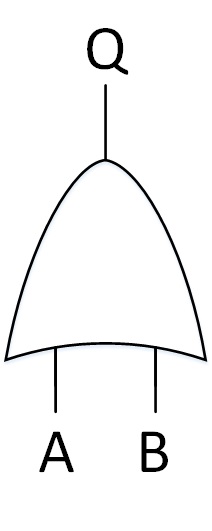}}}
\hfill\hfill\hfill \hfill \hfill  
\subfloat[PAND]{\label{fig:PAND}{\includegraphics[width=0.12\textwidth , height=3cm]{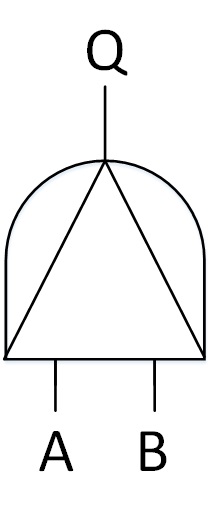}}}
\hfill\hfill\hfill \hfill \hfill 
\subfloat[FDEP]{\label{fig:FDEP}{\includegraphics[scale=0.2]{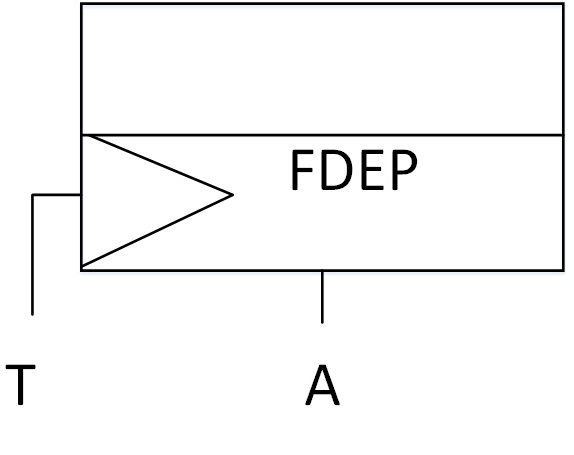}}}
\hfill \hfill \hfill \hfill 
\subfloat[Spare]{\label{fig:SPARE}{\includegraphics[scale=0.2]{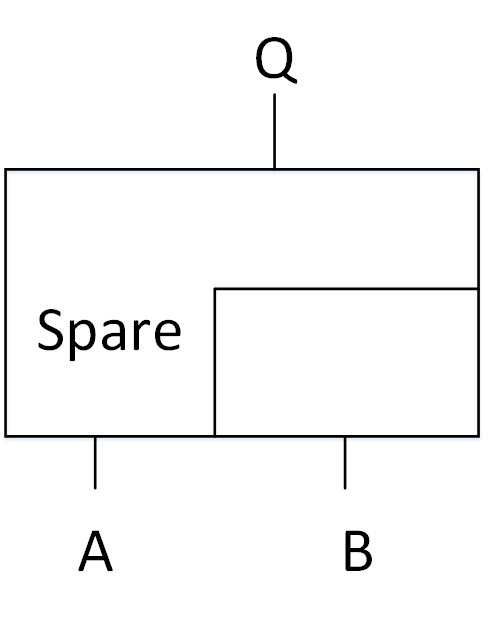}}}
%\vspace{-5pt}
\caption{Fault Tree Gates}
\vspace{-10pt}
\label{fig:DFT_Gates}
\end{figure}

Fig. \ref{fig:DFT_Gates} shows the main FT gates \cite{DFT-handbook}; dynamic gates as well as the static ones. \\
%\vspace{-35pt}
%\vspace{-10pt}

Although, the AND ($\cdot$) and OR (+) gates are considered as static operators or gates, their behavior can be represented using the time of occurrence of the input events. For example, the output event of an AND gate occurs if and only if all its input events occur. This implies that the output of the AND gate occurs with the occurrence of the last input event, which means that the time of occurrence of the output event equals the maximum time of occurrence of the input events. The OR gate is defined in a similar manner with the only difference that the output event occurs with the occurrence of the first input event, i.e., the minimum time of occurrence of the inputs:%\vspace{-8pt}

\begin{equation}
d(A \cdot B) = max (d(A)\ ,\ d(B))
\end{equation}
%\vspace{-16pt}
\begin{equation}
d(A + B) = min (d(A)\ ,\ d(B))
\end{equation}

\noindent We model the behavior of these gates in HOL as follows:%\vspace{-2pt}
\begin{defn}
\label{AND_def}
\emph{AND gate (operator) } \\
\vspace{1pt} {\texttt{$\vdash$ $\forall$ (A:extreal) B. D\_AND A B = max A B 
}}
\end{defn}%\vspace{-2pt}

\begin{defn}
\label{OR_def}
\emph{OR gate (operator) } \\
\vspace{1pt} {\texttt{$\vdash$ $\forall$ (A:extreal) B. D\_OR A B = min A B 
}}
\end{defn}\vspace{-2pt}

\noindent where \texttt{max} and \texttt{min} are functions that return the maximum and the minimum values of their arguments, respectively.

The Priority-AND (PAND) gate is a special case of the AND gate, where the output occurs when all the input events occur in sequence, conventionally from left to right. For the PAND gate, shown in Figure \ref{fig:PAND}, the output $Q$ occurs if $A$ and $B$ occur and $A$ occurs before or with $B$. The behavior of the PAND gate can be represented using the time of failure as:\vspace{-8pt}

\begin{equation}
d(Q) =  \begin{cases}d(B), &d(A) \leq d(B)\\ +\infty,  &d(A) > d(B) 
\end{cases} 
\end{equation}

\noindent The behavior of the PAND gate can be expressed using the temporal operators as:

\begin{equation}
\label{eq:PAND1}
Q =  B\cdot (A \unlhd B)
\end{equation}

\noindent We define the PAND in HOL as:\vspace{-2pt}

\begin{defn}
\label{PAND}
\emph{PAND gate} \\
\vspace{1pt} {\texttt{$\vdash$ $\forall$ (A:extreal) B.
PAND A B = if (A \(\leq\) B ) then B else PosInf
}}
\end{defn}\vspace{-2pt}

\noindent We verify in HOL that the PAND exhibits the behavior given in Equation \ref{eq:PAND1}:  \vspace{-2pt}
\begin{thm}
	\label{PAND2}
	\vspace{1pt} \texttt{$\vdash$ $\forall$ A B. PAND A B = D\_AND B (D\_INCLUSIVE\_BEFORE A B)
	}
\end{thm}

The Functional Dependency (FDEP) gate , shown in Figure \ref{fig:FDEP}, is used when there is a failure dependency between the input events or sub-trees, i.e., the occurrence of one input (or a sub-tree) can trigger the occurrence of other input events (or subtrees) in the fault tree. For example, in Figure \ref{fig:FDEP}, the occurrence of $T$ triggers the occurrence of $A$. This implies that $A$ occurs in two different ways: firstly, when $A$ occurs by itself, and secondly, when the trigger $T$ occurs. This means that the time of failure of $A_{T}$ (triggered $A$) equals the minimum time of occurrences of $T$ and $A$:
%\vspace{-10pt}
\begin{equation}
d(A_{T}) =  min(\ d(A),d(T) )
\end{equation}

\noindent We define the FDEP in HOL as:
\begin{defn}
\label{FDEP_def}
\emph{FDEP gate} \\
\vspace{1pt} {\texttt{$\vdash$ $\forall$ (A:extreal) T. FDEP A T = min A T 
}}
\end{defn}

\noindent where $T$ is the occurrence time of the trigger. We also verify in HOL that the FDEP is equivalent to an OR gate as follows: \vspace{-2pt}

\begin{thm}
	\label{FDEP2}
	\vspace{1pt} \texttt{$\vdash$ $\forall$ A T. FDEP A T = D\_OR A T
	}
\end{thm}

The spare gate, shown in Figure \ref{fig:SPARE}, represents a dynamic behavior that occurs in many real world systems, where we usually have a main part and some spare parts. The spare parts are utilized when the main part fails. The spare gate, shown in Figure \ref{fig:SPARE}, has a main input ($A$) and a spare input ($B$). After the failure of $A$, $B$ is activated. The output of the spare gate fails if both the main input and the spare fail. The spare gate can have several spare inputs, and the output fails after the failure of the main input and all the spares. The spare gate has three variants depending on the failure behavior of the spare part: the hot spare gate (HSP), the cold spare gate (CSP) and the warm spare gate (WSP). In the HSP, the probability of failure for the spare is the same in both the dormant and the active states. For the CSP, the spare part cannot fail unless it is activated. The WSP is the general case, where the spare part can fail in the dormant state as well as in the active state, but the failure distribution of the spare in its dormant state is different from the one in the active mode, and it is usually attenuated by a dormancy factor. In order to be able to distinguish between the different states of the spare input, two different variables are assigned to each state. For example, for the spare gate, shown in Figure \ref{fig:SPARE}, $B$ is represented using two variables; $B_{a}$ for the active state and $B_{d}$ for the dormant state.

The input events of the spare gate cannot occur at the same time if they are basic events. However, if these events are subtrees then they can occur at the same time. For a two input warm spare gate, with $A$ as the primary input and $B$ as the spare input, the output event occurs in two ways; firstly, if $A$ fails first then $B$, i.e., the spare part, is activated and then $B$ fails in its active state. The second way is when $B$ fails in its dormant state (inactive) then $A$ fails with no spare to replace it. For the general case where the input events can occur at the same time (if they are subtrees or depend on a common trigger), an additional option for the failure of the spare gate is added, where the two input events occur at the same time. This general warm spare gate can be described as:
\begin{equation}
Q = A.(B_{d} \lhd A) + B_{a}.(A \lhd B_{a}) + A \Delta B_{a} + A \Delta B_{d}
\end{equation}

\noindent We formalize the WSP in HOL as:
\begin{defn}
\label{WSP_SIMULT}
\emph{Warm Spare Gate} \\
\vspace{1pt} {\texttt{$\vdash$ $\forall$ A B\_a B\_d. WSP A B\_a B\_d =  D\_OR (D\_OR (D\_OR (D\_AND A(D\_BEFORE B\_d A))\\(D\_AND B\_a(D\_BEFORE A B\_a)))(D\_SIMULT A B\_a))(D\_SIMULT A B\_d) 
}}
\end{defn}
%
%If this definition is used as a CSP gate or as a warm spare gate (not as a HSP gate), then it should be considered that the spare part can either fail while it is dormant or active but not both. This is represented using the NEVER\_events in HOL as:
%
%\begin{defn}
%\label{NEVER_events}
%\emph{Never events} \\
%\vspace{1pt} {\texttt{$\vdash$ $\forall$ (B\_a:extreal) B\_d.\\
% NEVER\_events B\_a B\_d = (D\_AND B\_a B\_d = NEVER) 
%}}
%\end{defn}\vspace{-2pt}
%\noindent This definition will be exploited in Section \ref{Experiment results}, where spare gates are used with cold and warm spares.
The time of failure of the CSP gate with primary input $A$ and cold spare $B$ can be defined as:
\begin{equation}
d(Q) =  \begin{cases}d(B), &d(A) < d(B)\\ +\infty , &d(A) \geq d(B) \end{cases} 
\end{equation}

\noindent which means that the output event of the CSP occurs if the primary input fails and then the spare fails while in its active state. We define the CSP in HOL as:
\begin{defn}
\label{CSP_def}
\emph{Cold Spare Gate} \\
\vspace{1pt} {\texttt{$\vdash$ $\forall$ (A:extreal) B.  CSP A B = if (A < B) then B else PosInf
}}
\end{defn}

We verify in HOL that the CSP gate is a special case of WSP, where the spare part cannot fail in its dormant state. 
\begin{thm}
	\label{WSP_CSP}
	\vspace{1pt} \texttt{$\vdash$ $\forall$ A B\_a B\_d. \\ ALL\_DISTINCT [A; B\_a] \(\wedge\) COLD\_SPARE B\_d $\Rightarrow$ (WSP A B\_a B\_d = CSP A B\_a)
}
\end{thm}
\noindent where \texttt{ALL\_DISTINCT} ensures that $A$ and $B\_a$ are not equal, which means that they cannot fail at the same time, and \texttt{COLD\_SPARE B\_d} indicates that the spare $B$ is a cold spare, i.e., it cannot fail in its dormant mode ($B\_d$).

The spare part in the HSP has only one failure distribution, i.e., the dormant state and the active state are the same. The output of the HSP fails when both the primary and the spare fail, and the sequence of failure does not matter, as the spare part has only one failure distribution. The HSP is defined as:
\vspace{-4pt}
\begin{equation}
d(Q) =  max (d(A)\ ,\ d(B))
\end{equation}
\noindent where $A$ is the primary input and $B$ is the spare. We define this in HOL as:
\vspace{-4pt}
\begin{defn}
\label{HSP_def}
\emph{Hot Spare Gate} \\
\vspace{1pt} {\texttt{$\vdash$ $\forall$ (A:extreal) B. HSP A B = max A B
}}
\end{defn}

\noindent We verify in HOL that if both the dormant and the active states of the spare are equal, then the WSP is equivalent to the HSP:
\begin{thm}
	\label{WSP_HSP}
	\vspace{1pt} \texttt{$\vdash$ $\forall$ A B\_a B\_d. (B\_a = B\_d) $\Rightarrow$ (WSP A B\_a B\_d = HSP A B\_a)
}
\end{thm}
%\vspace{-8pt}
It is important to mention that more than one spare gate can share the same spare input. In this case, there is a possibility that one of the primary inputs is replaced by the spare, while the other input does not have a spare in case it fails. The outputs of the spare gates, shown in Figure \ref{fig:shared_spare}, are represented as follows (assuming that $A$, $B$ and $C$ are basic events): 
\begin{figure}[]
\centering
%\vspace{-30pt}
\includegraphics[scale=0.18]{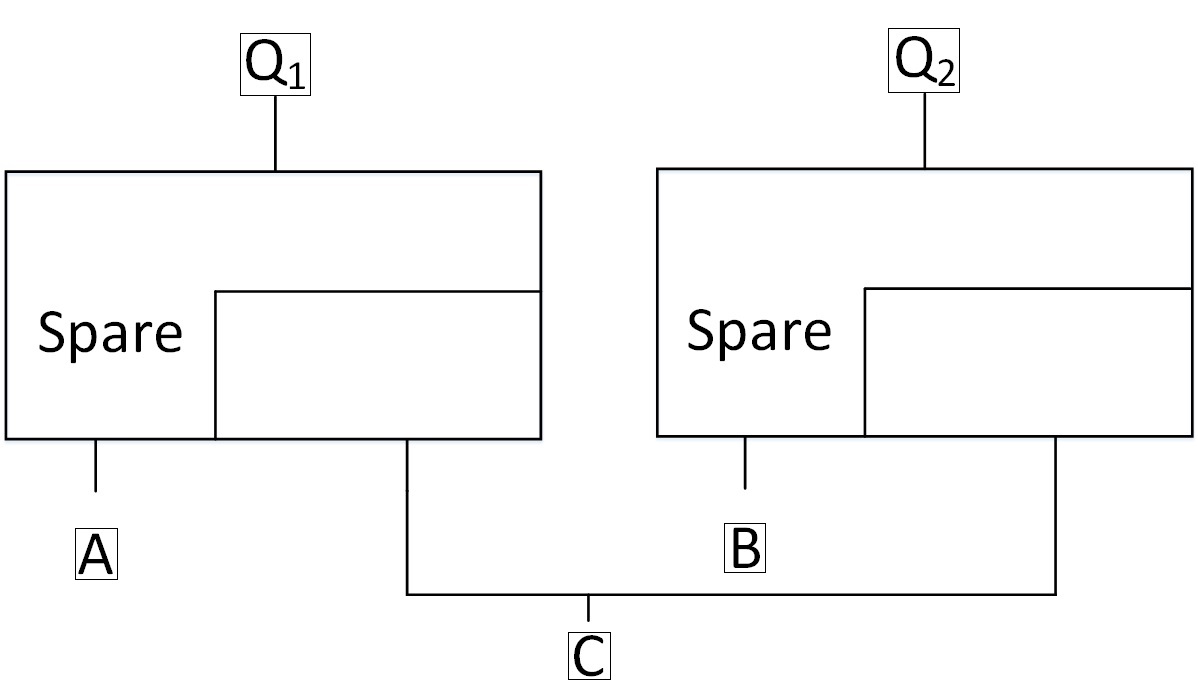}
%\vspace{-8pt}
\caption{Spare Gates with Shared Spare}
\label{fig:shared_spare}
\end{figure}
\begin{equation}
Q_{1} = A.(C_{d} \lhd A) + C_{a}.(A \lhd C_{a}) + A . (B \lhd A)
\end{equation}
\begin{equation}
Q_{2} = B.(C_{d} \lhd B) + C_{a}.(B \lhd C_{a}) + B . (A \lhd B)
\end{equation}

\noindent The last term in $Q_{1}$ indicates that if $B$ occurs before $A$, then the spare part $C$ is used by the second spare gate. This means that no spare is available for the first spare gate, which causes the failure of the output of the first spare gate if $A$ occurs. We formalize the output $Q_{1}$ of the first spare gate in HOL as:

\begin{defn}
\label{shared_spare}
\emph{Shared Spare} \\
\vspace{1pt}\noindent{\texttt{$\vdash$ $\forall$ A B C\_a C\_d. shared\_spare A B C\_a C\_d = D\_OR (D\_OR (D\_AND A(D\_BEFORE C\_d A))(D\_AND C\_a (D\_BEFORE A C\_a)))D\_AND  A(D\_BEFORE B A))  
}}
\end{defn}

We define a function in HOL called \texttt{Never\_events}, which ensures that its operands are mutual exclusive, i.e., only one of them can occur. We formalize it in HOL as:
\begin{defn}
\emph{Never events}\\
\vspace{1pt}\noindent{\texttt{$\vdash$ $\forall$ A B. NEVER\_events A B = (D\_AND A B = NEVER) 
}}
\end{defn}
\noindent This function is useful when we want to make sure that two events cannot happen together. For example, for a CSP gate, the spare part can only fail in one of its two states and not in both.

%\noindent where $A$ is the primary input, $C\_a$ and $C\_d$ represent the shared spare in its two modes and $B$ is the primary input of the second spare gate. 
%%that shares the same spare ($C$).
\section{Formal Verification of the Simplification Theorems}
\label{Simplification}
As with classical Boolean algebra, many simplification theorems also exist for DFT operators, which can be used to simplify the structure function of the DFT. We formally verified over 80 simplification theorems for the operators, defined in the previous section, including commutativity, associativity and idempotence of the AND, OR and Simultaneous operators, in addition to more complex theorems that include a combination of all operators. The verification process of these theorems was mainly based on the properties of extended real numbers, since the DFT operators are defined based on the time of failure of the events, which we choose to model using the \texttt{extreal} data type in HOL. During the verification process, each theorem was divided into several sub-goals based on the definition of the operators. Most of these sub-goals were automatically verified using automated tactics that utilize theorems from the \texttt{extreal} HOL theory. These simplification theorems can be classified into four groups depending on the operators involved in the simplification.

\subsection{Simplification Theorems using OR and AND}

These simplification theorems are similar to the OR and AND related Boolean algebra theorems, such as commutativity and associativity. Based on the theorems presented in \cite{Merle-thesis}, Table \ref{table:OR_AND_theorems} lists the formalization for these theorems, which we proved in HOL.

\begin{longtable}{ | p{9cm} | p{6cm} |}

\caption{Simplification Theorems for OR and AND\label{table:OR_AND_theorems}}\\
 
 %\hline
%\multicolumn{2}{| c |}{Begin of Table}\\
 \hline
 HOL Theorems & DFT Algebra Theorems\\
 \hline
 \endfirsthead
 
 %\hline
 %\multicolumn{2}{|c|}{Continuation of Table \ref{long}}\\
 \hline
 HOL Theorems &  DFT Algebra Theorems\\
 \hline
 \endhead
 
 \hline
 \endfoot
 
% \hline
 %\multicolumn{2}{| c |}{End of Table}\\
 %\hline\hline
 %\endlastfoot
 
\hline
	$\!\begin{aligned}[c]
	& \texttt{$\vdash$ $\forall$ A B. D\_OR A B = D\_OR B A  }\end{aligned}$& $\!\begin{aligned}[b]
	 A+B = B+A
	\end{aligned}$ \\	\hline
	
 	$\!\begin{aligned}[c]
	& \texttt{$\vdash$ $\forall$ A B. D\_AND A B = D\_AND B A  }\end{aligned}$& $\!\begin{aligned}[b]
	   A.B = B.A
	\end{aligned}$ \\	\hline
	
	$\!\begin{aligned}[c]
	& \texttt{$\vdash$ $\forall$ A B C. D\_OR A (D\_OR B C) =}\\
	&\texttt{D\_OR (D\_OR A B )C }\end{aligned}$& $\!\begin{aligned}[b]
	    A+(B+C) = (A+B)+C
	\end{aligned}$ \\	\hline

 	$\!\begin{aligned}[c]
	& \texttt{$\vdash$ $\forall$ A B C. D\_AND A (D\_AND B C) = }\\
	&\texttt{D\_AND (D\_AND A B ) C}\end{aligned}$& $\!\begin{aligned}[b]
	     A.(B.C) = (A.B).C
	\end{aligned}$ \\	\hline
	
  	$\!\begin{aligned}[c]
	& \texttt{$\vdash$ $\forall$ A . D\_OR A A = A }\end{aligned}$& $\!\begin{aligned}[b]
	       A+A = A
	\end{aligned}$ \\	\hline
	
 $\!\begin{aligned}[c]
	& \texttt{$\vdash$ $\forall$ A .  D\_AND A A = A }\end{aligned}$& $\!\begin{aligned}[b]
	       A.A = A
	\end{aligned}$ \\	\hline
	
  $\!\begin{aligned}[c]
	& \texttt{$\vdash$ $\forall$ A B C . D\_AND A (D\_OR B C) =}\\
	&\texttt{D\_OR (D\_AND A B)(D\_AND B C)}\end{aligned}$& $\!\begin{aligned}[b]
	       A.(B+C) = A.B + A.C
	\end{aligned}$ \\	\hline
	
$\!\begin{aligned}[c]
	& \texttt{$\vdash$ $\forall$ A. D\_OR A NEVER = A}\end{aligned}$& $\!\begin{aligned}[b]
	         A+NEVER = A
	\end{aligned}$ \\	\hline

 $\!\begin{aligned}[c]
	& \texttt{$\vdash$ $\forall$ A. D\_AND A ALWAYS = A}\end{aligned}$& $\!\begin{aligned}[b]
	        A.ALAWYS = A
	\end{aligned}$ \\	\hline

  $\!\begin{aligned}[c]
	& \texttt{$\vdash$ $\forall$ A.  D\_OR A ALWAYS = ALWAYS}\end{aligned}$& $\!\begin{aligned}[b]
	          A+ALAWYS = ALWAYS
	\end{aligned}$ \\	\hline

  $\!\begin{aligned}[c]
	& \texttt{$\vdash$ $\forall$ A.   D\_AND A NEVER = NEVER}\end{aligned}$& $\!\begin{aligned}[b]
	          A.NEVER = NEVER
	\end{aligned}$ \\	\hline

  $\!\begin{aligned}[c]
	& \texttt{$\vdash$ $\forall$ A B C. D\_OR A (D\_AND B C) =}\\
	&\texttt{D\_AND (D\_OR A B)(D\_OR A C)}\end{aligned}$& $\!\begin{aligned}[b]
	           A+(B.C) = (A+B).(A+C)
	\end{aligned}$ \\	\hline

  $\!\begin{aligned}[c]
	& \texttt{$\vdash$ $\forall$  A B. D\_OR A (D\_AND A B) = A}\end{aligned}$& $\!\begin{aligned}[b]
	           A+(A.B) = A
	\end{aligned}$ \\	\hline
	
  $\!\begin{aligned}[c]
	& \texttt{$\vdash$ $\forall$  A B. D\_AND A (D\_OR A B) = A}\end{aligned}$& $\!\begin{aligned}[b]
	           A.(A+B) = A
	\end{aligned}$ \\	\hline

  \end{longtable}

\subsection{Simplification Theorems using Before Operator}
ِAs with the AND and OR, several simplification theorems were introduced in \cite{Merle-thesis} to simplify expressions that include the Before operator. Our formalization of these theorems in HOL is given in Table \ref{table:Before_theorems}.

\begin{longtable}{ | p{9cm} | p{6cm} |}

\caption{Simplification Theorems for Before Operator\label{table:Before_theorems}}\\
 
 %\hline
%\multicolumn{2}{| c |}{Begin of Table}\\
 \hline
 HOL Theorems &  DFT Algebra Theorems\\
 \hline
 \endfirsthead
 
 %\hline
 %\multicolumn{2}{|c|}{Continuation of Table \ref{long}}\\
 \hline
 HOL Theorems &  DFT Algebra Theorems\\
 \hline
 \endhead
 
 \hline
 \endfoot
 
% \hline
 %\multicolumn{2}{| c |}{End of Table}\\
 %\hline\hline
 %\endlastfoot
   $\!\begin{aligned}[c]
	& \texttt{$\vdash$ $\forall$  A B.}\\
	&\texttt{D\_AND (D\_BEFORE A B) D\_BEFORE B A) = NEVER}\end{aligned}$& $\!\begin{aligned}[b]
	          (A \lhd B).(B \lhd A) = NEVER
	\end{aligned}$ \\	\hline

 $\!\begin{aligned}[c]
	& \texttt{$\vdash$ $\forall$  A B C.}\\
	&\texttt{D\_BEFORE A (D\_BEFORE B C) =}\\
	&\texttt{D\_OR (D\_BEFORE A B)(D\_AND (D\_AND A B)}\\
	&\texttt{(D\_OR (D\_BEFORE C B)(D\_SIMULT C B)))}\end{aligned}$& $\!\begin{aligned}[b]
	         A \lhd (B \lhd C) = (A \lhd B) + (A.B.\\((C \lhd B) + (C \Delta B)))
	\end{aligned}$ \\	\hline

 $\!\begin{aligned}[c]
	& \texttt{$\vdash$ $\forall$  A B C.}\\
	&\texttt{D\_BEFORE A (D\_BEFORE B C) = }\\
	&\texttt{D\_OR (D\_BEFORE A B)( D\_AND (D\_AND A B)}\\
	&\texttt{(D\_INCLUSIVE\_BEFORE C B))}\end{aligned}$& $\!\begin{aligned}[b]
	         A \lhd (B \lhd C) = (A \lhd B) + (A.B.\\((C \unlhd B)))
	\end{aligned}$ \\	\hline

 $\!\begin{aligned}[c]
	& \texttt{$\vdash$ $\forall$  A B C.}\\
	&\texttt{D\_BEFORE (D\_BEFORE A B) C =}\\
	&\texttt{D\_AND (D\_BEFORE A B)(D\_BEFORE A C)}\end{aligned}$& $\!\begin{aligned}[b]
(A \lhd B) \lhd C = (A \lhd B).(A \lhd C)
	\end{aligned}$ \\	\hline

 $\!\begin{aligned}[c]
	& \texttt{$\vdash$ $\forall$  A. D\_BEFORE NEVER A = NEVER}\end{aligned}$& $\!\begin{aligned}[b]
NEVER \lhd A = NEVER
	\end{aligned}$ \\	\hline

 $\!\begin{aligned}[c]
	& \texttt{$\vdash$ $\forall$  A. D\_BEFORE A NEVER = A}\end{aligned}$& $\!\begin{aligned}[b]
A \lhd NEVER = A
	\end{aligned}$ \\	\hline
	
	 $\!\begin{aligned}[c]
	& \texttt{$\vdash$ $\forall$  A. D\_BEFORE A A = NEVER}\end{aligned}$& $\!\begin{aligned}[b]
A \lhd A = NEVER
	\end{aligned}$ \\	\hline

 $\!\begin{aligned}[c]
	& \texttt{$\vdash$ $\forall$  A B C.}\\
	&\texttt{D\_BEFORE A (D\_OR B C) =}\\
	&\texttt{D\_AND (D\_BEFORE A B)(D\_BEFORE A C)}\end{aligned}$& $\!\begin{aligned}[b]
A \lhd (B+C) = (A \lhd B).( A \lhd C)
	\end{aligned}$ \\	\hline

 $\!\begin{aligned}[c]
	& \texttt{$\vdash$ $\forall$  A B C.}\\
	&\texttt{D\_BEFORE A (D\_AND B C) =}\\
	&\texttt{D\_OR (D\_BEFORE A B)(D\_BEFORE A C)}\end{aligned}$& $\!\begin{aligned}[b]
A \lhd (B.C) = (A \lhd B)+( A \lhd C)
	\end{aligned}$ \\	\hline
	
 $\!\begin{aligned}[c]
	& \texttt{$\vdash$ $\forall$  A B C.}\\
	&\texttt{D\_BEFORE A (D\_SIMULT B C) = D\_OR (D\_OR}\\
	&\texttt{(D\_OR (D\_AND A (D\_BEFORE B C))}\\
	&\texttt{(D\_AND A (D\_BEFORE C B)))}\\
	&\texttt{(D\_BEFORE A B))(D\_BEFORE A C)}\end{aligned}$& $\!\begin{aligned}[b]
A \lhd (B \Delta C) = (A.(B \lhd C))+\\(A.(C \lhd B))+(A \lhd B)+\\(A \lhd C)
	\end{aligned}$ \\	\hline
	
$\!\begin{aligned}[c]
	& \texttt{$\vdash$ $\forall$  A B C.}\\
	&\texttt{D\_BEFORE A (D\_INCLUSIVE\_BEFORE B C) =}\\
	&\texttt{D\_OR (D\_BEFORE A B)(D\_AND}\\
	&\texttt{(D\_AND A B)(D\_BEFORE C B))}\end{aligned}$& $\!\begin{aligned}[b]
A \lhd (B \unlhd C) = (A \lhd B)+\\(A.B.(C \lhd B))
	\end{aligned}$ \\	\hline

%3.27

$\!\begin{aligned}[c]
	& \texttt{$\vdash$ $\forall$  A B C.}\\
	&\texttt{D\_BEFORE (D\_OR A B) C =}\\
	&\texttt{D\_OR (D\_BEFORE A C)(D\_BEFORE B C)}\end{aligned}$& $\!\begin{aligned}[b]
(A+B) \lhd C = (A \lhd C)+\\(B \lhd C)
	\end{aligned}$ \\	\hline

%3.28

$\!\begin{aligned}[c]
	& \texttt{$\vdash$ $\forall$  A B C.}\\
	&\texttt{D\_BEFORE (D\_AND A B) C =}\\
	&\texttt{D\_AND (D\_BEFORE A C)(D\_BEFORE B C)}\end{aligned}$& $\!\begin{aligned}[b]
(A.B) \lhd C = (A \lhd C).(B \lhd C)
	\end{aligned}$ \\	\hline

%3.29

$\!\begin{aligned}[c]
	& \texttt{$\vdash$ $\forall$  A B C.}\\
	&\texttt{D\_BEFORE (D\_SIMULT A B) C =}\\
	&\texttt{D\_AND (D\_SIMULT A B)(D\_BEFORE A C)}\end{aligned}$& $\!\begin{aligned}[b]
(A \Delta B) \lhd C = (A \Delta B).(A \lhd C)
	\end{aligned}$ \\	\hline

%3.29
$\!\begin{aligned}[c]
	& \texttt{$\vdash$ $\forall$  A B C.}\\
	&\texttt{D\_BEFORE (D\_SIMULT A B) C =}\\
	&\texttt{D\_AND (D\_SIMULT A B)(D\_BEFORE B C)}\end{aligned}$& $\!\begin{aligned}[b]
(A \Delta B) \lhd C = (A \Delta B).(B \lhd C)
	\end{aligned}$ \\	\hline

%3.29
$\!\begin{aligned}[c]
	& \texttt{$\vdash$ $\forall$  A B C.}\\
	&\texttt{D\_BEFORE (D\_SIMULT A B) C =}\\
	&\texttt{D\_SIMULT (D\_BEFORE A C)(D\_BEFORE B C)}\end{aligned}$& $\!\begin{aligned}[b]
(A \Delta B) \lhd C = (A \lhd C) \Delta \\ (B \lhd C)
	\end{aligned}$ \\	\hline

%3.30

$\!\begin{aligned}[c]
	& \texttt{$\vdash$ $\forall$  A B C.}\\
	&\texttt{D\_BEFORE (D\_INCLUSIVE\_BEFORE A B) C =}\\
	&\texttt{D\_AND (D\_INCLUSIVE\_BEFORE A B)}\\
	&\texttt{(D\_BEFORE A C)}\end{aligned}$& $\!\begin{aligned}[b]
(A \unlhd B) \lhd C = (A \unlhd B).(A \lhd C)
	\end{aligned}$ \\	\hline

%3.31

$\!\begin{aligned}[c]
	& \texttt{$\vdash$ $\forall$  A B. D\_OR A(D\_BEFORE A B) = A}\end{aligned}$& $\!\begin{aligned}[b]
A+(A \lhd B) = A
	\end{aligned}$ \\	\hline

%3.32

$\!\begin{aligned}[c]
	& \texttt{$\vdash$ $\forall$  A B.}\\
	&\texttt{D\_OR (D\_BEFORE A B)B = D\_OR A B}\end{aligned}$& $\!\begin{aligned}[b]
(A \lhd B)+B = A+B
	\end{aligned}$ \\	\hline

%3.33
$\!\begin{aligned}[c]
	& \texttt{$\vdash$ $\forall$  A B.}\\
	&\texttt{D\_AND A(D\_BEFORE A B) =  D\_BEFORE A B}\end{aligned}$& $\!\begin{aligned}[b]
A.(A \lhd B) = A \lhd B
	\end{aligned}$ \\	\hline

%3.34
$\!\begin{aligned}[c]
	& \texttt{$\vdash$ $\forall$  A B C.}\\
	&\texttt{D\_AND (D\_AND (D\_BEFORE A B)}\\
	&\texttt{(D\_BEFORE B C))(D\_BEFORE A C) =}\\
	&\texttt{D\_AND (D\_BEFORE A B)(D\_BEFORE B C)}\end{aligned}$& $\!\begin{aligned}[b]
(A \lhd B).(B \lhd C).(A \lhd C) = \\(A \lhd B).(B \lhd C)
	\end{aligned}$ \\	\hline
	
\end{longtable}

\subsection{Simplification Theorems using Simultaneous Operator}
Table \ref{table:SIMULT_theorems} shows the simplification theorems which can be used with the Simultaneous operator along with their formalizations in HOL.

\begin{longtable}{ | p{9cm} | p{6cm} |}

\caption{Simplification Theorems for Simultaneous Operator\label{table:SIMULT_theorems}}\\
 
 %\hline
%\multicolumn{2}{| c |}{Begin of Table}\\
 \hline
 HOL Theorems &  DFT Algebra Theorems\\
 \hline
 \endfirsthead
 
 %\hline
 %\multicolumn{2}{|c|}{Continuation of Table \ref{long}}\\
 \hline
 HOL Theorems &  DFT Algebra Theorems\\
 \hline
 \endhead
 
 \hline
 \endfoot
 
% \hline
 %\multicolumn{2}{| c |}{End of Table}\\
 %\hline\hline
 %\endlastfoot
%3.35
$\!\begin{aligned}[c]
	& \texttt{$\vdash$ $\forall$  A B. D\_SIMULT A B = D\_SIMULT B A}\end{aligned}$& $\!\begin{aligned}[b]
  A \Delta B = B \Delta A
	\end{aligned}$ \\	\hline

%3.36

$\!\begin{aligned}[c]
	& \texttt{$\vdash$ $\forall$  A B C.}\\
	&\texttt{D\_SIMULT A (D\_SIMULT B C) =}\\
	&\texttt{D\_SIMULT (D\_SIMULT A B) C}\end{aligned}$& $\!\begin{aligned}[b]
A \Delta (B \Delta C) = (A \Delta B) \Delta C
	\end{aligned}$ \\	\hline

%3.36
$\!\begin{aligned}[c]
	& \texttt{$\vdash$ $\forall$  A B C.}\\
	&\texttt{D\_SIMULT A (D\_SIMULT B C) =}\\
	&\texttt{D\_AND (D\_SIMULT A B)(D\_SIMULT B C)}\end{aligned}$& $\!\begin{aligned}[b]
A \Delta (B \Delta C) = (A \Delta B).(B \Delta C)
	\end{aligned}$ \\	\hline

%3.36
$\!\begin{aligned}[c]
	& \texttt{$\vdash$ $\forall$  A B C.}\\
	&\texttt{D\_SIMULT A (D\_SIMULT B C) =}\\
	&\texttt{D\_AND (D\_SIMULT A C)(D\_SIMULT C B)}\end{aligned}$& $\!\begin{aligned}[b]
A \Delta (B \Delta C) = (A \Delta C).(C \Delta B)
	\end{aligned}$ \\	\hline

%3.37
$\!\begin{aligned}[c]
	& \texttt{$\vdash$ $\forall$  A. D\_SIMULT A NEVER = NEVER}\end{aligned}$& $\!\begin{aligned}[b]
A \Delta NEVER = NEVER
	\end{aligned}$ \\	\hline

%3.38
$\!\begin{aligned}[c]
	& \texttt{$\vdash$ $\forall$  A. D\_SIMULT A A = A}\end{aligned}$& $\!\begin{aligned}[b]
A \Delta A = A
	\end{aligned}$ \\	\hline

%3.39
$\!\begin{aligned}[c]
	& \texttt{$\vdash$ $\forall$  A B C.}\\
	&\texttt{D\_SIMULT A (D\_OR B C) = D\_OR (D\_OR}\\
	&\texttt{(D\_AND (D\_SIMULT A B)(D\_SIMULT B C))}\\
	&\texttt{(D\_AND (D\_SIMULT A B)(D\_BEFORE B C)))}\\
	&\texttt{(D\_AND (D\_SIMULT A C)(D\_BEFORE C B))}\end{aligned}$& $\!\begin{aligned}[b]
A \Delta (B + C) = (A \Delta B).(B \Delta C)\\+(A \Delta B).(B \lhd C)\\+(A \Delta C).(C \lhd B)
	\end{aligned}$ \\	\hline

%3.39
$\!\begin{aligned}[c]
	& \texttt{$\vdash$ $\forall$  A B C.}\\
	&\texttt{D\_SIMULT A (D\_OR B C) = D\_OR}\\
	&\texttt{(D\_AND (D\_SIMULT A B)}\\
	&\texttt{(D\_INCLUSIVE\_BEFORE B C))(D\_AND}\\
	&\texttt{(D\_SIMULT A C)(D\_INCLUSIVE\_BEFORE C B))}\end{aligned}$& $\!\begin{aligned}[b]
A \Delta (B + C) = (A \Delta B).(B \unlhd C)+\\(A \Delta C).(C \unlhd B)
	\end{aligned}$ \\	\hline

%3.40
$\!\begin{aligned}[c]
	& \texttt{$\vdash$ $\forall$  A B C.}\\
	&\texttt{D\_SIMULT A (D\_AND B C) = D\_OR}\\
	&\texttt{(D\_OR (D\_AND (D\_SIMULT A B)}\\
	&\texttt{(D\_SIMULT B C))(D\_AND (D\_SIMULT A B)}\\
	&\texttt{(D\_BEFORE C B)))(D\_AND}\\
	&\texttt{(D\_SIMULT A C)(D\_BEFORE B C))}\end{aligned}$& $\!\begin{aligned}[b]
A \Delta (B . C) = (A \Delta B).(B \Delta C)\\+(A \Delta B).(C \lhd B)\\+(A \Delta C).(B \lhd C)
	\end{aligned}$ \\	\hline
	
%3.40
$\!\begin{aligned}[c]
	& \texttt{$\vdash$ $\forall$  A B C.}\\
	&\texttt{D\_SIMULT A (D\_AND B C) =}\\
	&\texttt{D\_OR (D\_AND (D\_SIMULT A B)}\\
	&\texttt{(D\_INCLUSIVE\_BEFORE C B))}\\
	&\texttt{(D\_AND (D\_SIMULT A C)}\\
	&\texttt{(D\_INCLUSIVE\_BEFORE B C))}\end{aligned}$& $\!\begin{aligned}[b]
A \Delta (B . C) = (A \Delta B).(C \unlhd B)\\+(A \Delta C).(B \unlhd C)
	\end{aligned}$ \\	\hline

%3.41
$\!\begin{aligned}[c]
	& \texttt{$\vdash$ $\forall$  A B C.}\\
	&\texttt{D\_SIMULT A (D\_BEFORE B C) =}\\
	&\texttt{D\_AND (D\_SIMULT A B)(D\_BEFORE B C)}\end{aligned}$& $\!\begin{aligned}[b]
A \Delta (B \lhd C) = (A \Delta B).(B \lhd C)
	\end{aligned}$ \\	\hline

%3.42

$\!\begin{aligned}[c]
	& \texttt{$\vdash$ $\forall$  A B C.}\\
	&\texttt{D\_SIMULT A (D\_INCLUSIVE\_BEFORE B C) =}\\
	&\texttt{D\_AND (D\_SIMULT A B)}\\
	&\texttt{(D\_INCLUSIVE\_BEFORE B C)}\end{aligned}$& $\!\begin{aligned}[b]
A \Delta (B \unlhd C) = (A \Delta B).(B \unlhd C)
	\end{aligned}$ \\	\hline

%3.43

$\!\begin{aligned}[c]
	& \texttt{$\vdash$ $\forall$  A B.}\\
	&\texttt{D\_OR A (D\_SIMULT A B) = A}\end{aligned}$& $\!\begin{aligned}[b]
A+(A \Delta B) = A
	\end{aligned}$ \\	\hline

%3.44

$\!\begin{aligned}[c]
	& \texttt{$\vdash$ $\forall$  A B.}\\
	&\texttt{D\_AND A (D\_SIMULT A B) = D\_SIMULT A B}\end{aligned}$& $\!\begin{aligned}[b]
A.(A \Delta B)=A \Delta B
	\end{aligned}$ \\	\hline

%3.45

$\!\begin{aligned}[c]
	& \texttt{$\vdash$ $\forall$  A B C.}\\
	&\texttt{D\_AND (D\_AND (D\_SIMULT A B)}\\
	&\texttt{(D\_SIMULT B C))(D\_SIMULT A C) =}\\
	&\texttt{D\_AND (D\_SIMULT A B)(D\_SIMULT B C)}\end{aligned}$& $\!\begin{aligned}[b]
(A \Delta B).(B \Delta C).(A \Delta C)=\\(A \Delta B).(B \Delta C)
	\end{aligned}$ \\	\hline

\end{longtable}

\subsection{Simplification Theorems using Inclusive Before Operator}

Table \ref{table:Inclusive_Before_theorems} shows the HOL verified formalization of the theorems that can be used with the Inclusive Before operator.

\begin{longtable}{ | p{9cm} | p{6cm} |}

\caption{Simplification Theorems for Inclusive Before Operator\label{table:Inclusive_Before_theorems}}\\
 
 \hline
 HOL Theorems &  DFT Algebra Theorems\\
 \hline
 \endfirsthead

 \hline 
 HOL Theorems &  DFT Algebra Theorems\\
 \hline
 \endhead
 
 \hline
 \endfoot
%3.46

$\!\begin{aligned}[c]
	& \texttt{$\vdash$ $\forall$  A B.}\\
	&\texttt{D\_AND (D\_INCLUSIVE\_BEFORE A B)}\\
	&\texttt{(D\_INCLUSIVE\_BEFORE B A) = D\_SIMULT A B}\end{aligned}$& $\!\begin{aligned}[b]
 (A \unlhd B).(B \unlhd A) = A \Delta B
	\end{aligned}$ \\	\hline

%3.47

$\!\begin{aligned}[c]
	& \texttt{$\vdash$ $\forall$  A B C.}\\
	&\texttt{D\_INCLUSIVE\_BEFORE A}\\
	&\texttt{(D\_INCLUSIVE\_BEFORE B C) =}\\
	&\texttt{D\_OR (D\_OR (D\_BEFORE A B)}\\
	&\texttt{(D\_AND (D\_AND A B)(D\_BEFORE C B)))}\\
	&\texttt{(D\_AND (D\_SIMULT A B)}\\
	&\texttt{(D\_INCLUSIVE\_BEFORE B C))}\end{aligned}$& $\!\begin{aligned}[b]
A \unlhd (B \unlhd C) = (A \lhd B)\\+(A.B.(C \lhd B))\\+(A \Delta B).(B \unlhd C)
	\end{aligned}$ \\	\hline

%3.48

$\!\begin{aligned}[c]
	& \texttt{$\vdash$ $\forall$  A B C.}\\
	&\texttt{D\_INCLUSIVE\_BEFORE}\\
	&\texttt{(D\_INCLUSIVE\_BEFORE A B) C =}\\
	&\texttt{D\_AND (D\_INCLUSIVE\_BEFORE A B)}\\
	&\texttt{(D\_INCLUSIVE\_BEFORE A C)}\end{aligned}$& $\!\begin{aligned}[b]
(A \unlhd B) \unlhd C = (A \unlhd B).(A \unlhd C)
	\end{aligned}$ \\	\hline
	
%3.49
$\!\begin{aligned}[c]
	& \texttt{$\vdash$ $\forall$  A.}\\
	&\texttt{D\_INCLUSIVE\_BEFORE NEVER A = NEVER}\end{aligned}$& $\!\begin{aligned}[b]
NEVER \unlhd A = NEVER
	\end{aligned}$ \\	\hline

%3.50

$\!\begin{aligned}[c]
	& \texttt{$\vdash$ $\forall$  A.}\\
	&\texttt{D\_INCLUSIVE\_BEFORE A NEVER = A }\end{aligned}$& $\!\begin{aligned}[b]
 A \unlhd NEVER = A
	\end{aligned}$ \\	\hline

%3.51
$\!\begin{aligned}[c]
	& \texttt{$\vdash$ $\forall$  A.}\\
	&\texttt{D\_INCLUSIVE\_BEFORE A A = A }\end{aligned}$& $\!\begin{aligned}[b]
 A \unlhd A = A
	\end{aligned}$ \\	\hline

%3.52
$\!\begin{aligned}[c]
	& \texttt{$\vdash$ $\forall$  A B C.}\\
	&\texttt{D\_INCLUSIVE\_BEFORE A (D\_OR B C) =}\\
	&\texttt{D\_AND (D\_INCLUSIVE\_BEFORE A B)}\\
	&\texttt{(D\_INCLUSIVE\_BEFORE A C)}\end{aligned}$& $\!\begin{aligned}[b]
 A \unlhd (B+C)=(A \unlhd B).(A \unlhd C)
	\end{aligned}$ \\	\hline

%3.53

$\!\begin{aligned}[c]
	& \texttt{$\vdash$ $\forall$  A B C.}\\
	&\texttt{D\_INCLUSIVE\_BEFORE A (D\_AND B C) =}\\
	&\texttt{D\_OR (D\_INCLUSIVE\_BEFORE A B)}\\
	&\texttt{(D\_INCLUSIVE\_BEFORE A C)}\end{aligned}$& $\!\begin{aligned}[b]
 A \unlhd (B.C)=(A \unlhd B)+(A \unlhd C)
	\end{aligned}$ \\	\hline

%3.54

$\!\begin{aligned}[c]
	& \texttt{$\vdash$ $\forall$  A B C.}\\
	&\texttt{D\_INCLUSIVE\_BEFORE A (D\_BEFORE B C) =}\\
	&\texttt{D\_OR(D\_OR(D\_BEFORE A B)(D\_AND (D\_AND A B)}\\
	&\texttt{(D\_INCLUSIVE\_BEFORE C B)))}\\
	&\texttt{(D\_AND (D\_SIMULT A B)(D\_BEFORE B C))}\end{aligned}$& $\!\begin{aligned}[b]
 A \unlhd (B \lhd C)= (A \lhd B)\\+(A.B.(C \unlhd B)\\+(A \Delta B).(B \lhd C)
	\end{aligned}$ \\	\hline

%3.55

$\!\begin{aligned}[c]
	& \texttt{$\vdash$ $\forall$  A B C.}\\
	&\texttt{D\_INCLUSIVE\_BEFORE A (D\_SIMULT B C) =}\\
	&\texttt{D\_OR(D\_OR(D\_OR(D\_OR(D\_AND A}\\
	&\texttt{(D\_BEFORE B C))(D\_AND A}\\
	&\texttt{(D\_BEFORE C B)))(D\_BEFORE A B))}\\
	&\texttt{(D\_BEFORE A C))(D\_AND}\\
	&\texttt{(D\_SIMULT A B)(D\_SIMULT B C))}\end{aligned}$& $\!\begin{aligned}[c]
 A \unlhd (B \Delta C)=(A.(B \lhd C))\\+(A.(C \lhd B))\\+(A \lhd B)+(A \lhd C)\\+(A \Delta B).(B \Delta C)
	\end{aligned}$ \\	\hline

%3.56
$\!\begin{aligned}[c]
	& \texttt{$\vdash$ $\forall$  A B C.}\\
	&\texttt{D\_INCLUSIVE\_BEFORE (D\_OR A B) C =}\\
	&\texttt{D\_OR (D\_INCLUSIVE\_BEFORE A C)}\\
	&\texttt{(D\_INCLUSIVE\_BEFORE B C)}\end{aligned}$& $\!\begin{aligned}[b]
(A+B) \unlhd C = (A \unlhd C)\\+(B \unlhd C)
	\end{aligned}$ \\	\hline

%3.57
	$\!\begin{aligned}[c]
	& \texttt{$\vdash$ $\forall$ A B C. }\\
	&\texttt{D\_INCLUSIVE\_BEFORE (D\_AND A B) C=}\\ &\texttt{D\_AND (D\_INCLUSIVE\_BEFORE A C)}\\
	&\texttt{(D\_INCLUSIVE\_BEFORE B C)}\end{aligned}$& 
	$\!\begin{aligned}[b]
	(A.B) \unlhd C = (A \unlhd C).(B \unlhd C)
	\end{aligned}$\\
	\hline

%
%
%3.58
	$\!\begin{aligned}[c]
	& \texttt{$\vdash$ $\forall$ A B C. }\\
	&\texttt{D\_INCLUSIVE\_BEFORE (D\_SIMULT A B) C =}\\ 
	&\texttt{D\_AND (D\_SIMULT A B)}\\
	&\texttt{(D\_INCLUSIVE\_BEFORE A C)}\end{aligned}$& 
	$\!\begin{aligned}[b]
	(A \Delta B) \unlhd C = (A \Delta B).(A \unlhd C)
	\end{aligned}$\\
	\hline

%3.58
	$\!\begin{aligned}[c]
	& \texttt{$\vdash$ $\forall$ A B C. }\\
	&\texttt{D\_INCLUSIVE\_BEFORE (D\_SIMULT A B) C =}\\ 
	&\texttt{D\_AND (D\_SIMULT A B)}\\
	&\texttt{(D\_INCLUSIVE\_BEFORE B C)}\end{aligned}$& 
	$\!\begin{aligned}[b]
	(A \Delta B) \unlhd C = (A \Delta B).(B \unlhd C)
	\end{aligned}$\\
	\hline

%3.58
	$\!\begin{aligned}[c]
	& \texttt{$\vdash$ $\forall$ A B C. }\\
	&\texttt{D\_INCLUSIVE\_BEFORE (D\_SIMULT A B) C =}\\
	&\texttt{D\_SIMULT (D\_INCLUSIVE\_BEFORE A C)}\\
	&\texttt{(D\_INCLUSIVE\_BEFORE B C)}\end{aligned}$& 
	$\!\begin{aligned}[b]
	(A \Delta B) \unlhd C = (A \unlhd C)\\ \Delta (B \unlhd C)
	\end{aligned}$\\
	\hline

%3.59
	$\!\begin{aligned}[c]
	& \texttt{$\vdash$ $\forall$ A B C. }\\
	&\texttt{D\_INCLUSIVE\_BEFORE (D\_BEFORE A B) C =}\\ &\texttt{D\_AND (D\_BEFORE A B)}\\
	&\texttt{(D\_INCLUSIVE\_BEFORE A C)}\end{aligned}$& 
	$\!\begin{aligned}[b]
	(A \lhd B) \unlhd C = (A \lhd B).(A \unlhd C)
	\end{aligned}$\\
	\hline
%3.60
	$\!\begin{aligned}[c]
	& \texttt{$\vdash$ $\forall$ A B. }\\
	&\texttt{D\_OR A (D\_INCLUSIVE\_BEFORE A B) = A}\end{aligned}$& 
	$\!\begin{aligned}[b]
	A+ (A \unlhd B) = A
	\end{aligned}$\\
	\hline
	
%3.61
	$\!\begin{aligned}[c]
	& \texttt{$\vdash$ $\forall$ A B. }\\
	&\texttt{D\_OR B (D\_INCLUSIVE\_BEFORE A B) = D\_OR A B}\end{aligned}$& 
	$\!\begin{aligned}[b]
	B+ (A \unlhd B) = A+B
	\end{aligned}$\\
	\hline	
	
	%3.62
	$\!\begin{aligned}[c]
	& \texttt{$\vdash$ $\forall$ A B. }\\
	&\texttt{D\_AND A (D\_INCLUSIVE\_BEFORE A B) =}\\
	&\texttt{D\_INCLUSIVE\_BEFORE A B}\end{aligned}$& 
	$\!\begin{aligned}[b]
	A.(A \unlhd B) = A \unlhd B
	\end{aligned}$\\
	\hline	
	
		%3.63
	$\!\begin{aligned}[c]
	& \texttt{$\vdash$ $\forall$ A B. }\\
	&\texttt{D\_OR (D\_INCLUSIVE\_BEFORE A B)}\\
	&\texttt {(D\_INCLUSIVE\_BEFORE B A) = D\_OR A B}\end{aligned}$& 
	$\!\begin{aligned}[b]
	(A \unlhd B)+(B \unlhd A) = A+B
	\end{aligned}$\\
	\hline	
	%3.64
	$\!\begin{aligned}[c]
	& \texttt{$\vdash$ $\forall$ A B. }\\
	&\texttt{D\_OR (D\_AND A (D\_INCLUSIVE\_BEFORE A B))}\\
	&\texttt {(D\_AND B (D\_INCLUSIVE\_BEFORE B A)) =}\\
	&\texttt{D\_AND A B}\end{aligned}$& 
	$\!\begin{aligned}[b]
	(A.(B \unlhd A))+(B.(A \unlhd B)) \\= A.B
	\end{aligned}$\\
	\hline	
	
		%3.65
	$\!\begin{aligned}[c]
	& \texttt{$\vdash$ $\forall$ A B. }\\
	&\texttt{D\_OR (D\_INCLUSIVE\_BEFORE A B)}\\
	&\texttt{(D\_AND A (D\_INCLUSIVE\_BEFORE B A)) = A}\end{aligned}$& 
	$\!\begin{aligned}[b]
	(A \unlhd B)+(A.(B \unlhd B)) = A
	\end{aligned}$\\
	\hline	
	%3.66
	$\!\begin{aligned}[c]
	& \texttt{$\vdash$ $\forall$ A B C. }\\
	&\texttt{D\_AND (D\_AND (D\_INCLUSIVE\_BEFORE A B)}\\
	&\texttt{(D\_INCLUSIVE\_BEFORE B C))}\\
	&\texttt{(D\_INCLUSIVE\_BEFORE A C)) = }\\
	&\texttt{D\_AND (D\_INCLUSIVE\_BEFORE A B)}\\
	&\texttt{(D\_INCLUSIVE\_BEFORE B C)}\end{aligned}$& 
	$\!\begin{aligned}[b]
	(A\unlhd B).(B \unlhd C).(A \unlhd C)\\=(A\unlhd B).(B\unlhd C)
	\end{aligned}$\\
	\hline

\end{longtable}

\subsection{Simplification Theorems for Combinations of Operators}

Table \ref{table:all_theorems} shows our formalization in HOL for some simplification theorems from \cite{Merle-thesis}, which can be used to simplify expressions involving combinations of operators.

\begin{longtable}{ | p{9cm} | p{6cm} |}	
	\caption{Simplification Theorems for Combinations of Operators
	\label{table:all_theorems}}\\	
	\hline
	HOL Theorems &  DFT Algebra Theorems\\
	\hline
	\endfirsthead	
	\hline
	HOL Theorems &  DFT Algebra Theorems\\
	\hline
	\endhead	
	\hline
	\endfoot	
	
	%3.67
	$\!\begin{aligned}[c]
	& \texttt{$\vdash$ $\forall$ A B. }\\
	&\texttt{D\_OR (D\_INCLUSIVE\_BEFORE A B)}\\
	&\texttt {(D\_BEFORE A B) = D\_INCLUSIVE\_BEFORE A B}\end{aligned}$& 
	$\!\begin{aligned}[b]
	(A\unlhd B)+(A \lhd B) = A \unlhd B
	\end{aligned}$\\
	\hline	
	
%3.68
	$\!\begin{aligned}[c]
	& \texttt{$\vdash$ $\forall$ A B. }\\
	&\texttt{D\_OR (D\_INCLUSIVE\_BEFORE A B)}\\
	&\texttt {(D\_SIMULT A B) = D\_INCLUSIVE\_BEFORE A B}\end{aligned}$& 
	$\!\begin{aligned}[b]
	(A\unlhd B)+(A \Delta B) = A \unlhd B
	\end{aligned}$\\
	\hline

%3.69
	$\!\begin{aligned}[c]
	& \texttt{$\vdash$ $\forall$ A B. }\\
	&\texttt{D\_AND (D\_BEFORE A B)(D\_SIMULT A B) = NEVER}\end{aligned}$& 
	$\!\begin{aligned}[b]
	(A \lhd B).(A \Delta B) = NEVER
	\end{aligned}$\\
	\hline

%3.70
	$\!\begin{aligned}[c]
	& \texttt{$\vdash$ $\forall$ A B C. }\\
	&\texttt{D\_AND (D\_BEFORE A B)(D\_SIMULT B C)=}\\
	&\texttt {D\_AND (D\_BEFORE A C)(D\_SIMULT B C)}\end{aligned}$& 
	$\!\begin{aligned}[c]
	(A \lhd B).(B \Delta C)=\\(A \lhd C).(B \Delta C)
	\end{aligned}$\\
	\hline		
%3.71
	$\!\begin{aligned}[c]
	& \texttt{$\vdash$ $\forall$ A B. }\\
	&\texttt{D\_AND (D\_INCLUSIVE\_BEFORE A B)}\\
	&\texttt {(D\_BEFORE A B)) = D\_BEFORE A B}\end{aligned}$& 
	$\!\begin{aligned}[b]
	(A\unlhd B).(A \lhd B) = A \lhd B
	\end{aligned}$\\
	\hline	
	
%3.72
	$\!\begin{aligned}[c]
	& \texttt{$\vdash$ $\forall$ A B. }\\
	&\texttt{D\_AND (D\_BEFORE A B)}\\
	&\texttt {(D\_INCLUSIVE\_BEFORE B A) = NEVER}\end{aligned}$& 
	$\!\begin{aligned}[b]
	(A\lhd B).(B \unlhd A) = NEVER
	\end{aligned}$\\
	\hline

%3.73
	$\!\begin{aligned}[c]
	& \texttt{$\vdash$ $\forall$ A B. }\\
	&\texttt{D\_AND (D\_INCLUSIVE\_BEFORE A B)}\\
	&\texttt {(D\_SIMULT A B) = D\_SIMULT A B}\end{aligned}$& 
	$\!\begin{aligned}[b]
	(A\unlhd B).(A \Delta B) = A \Delta B
	\end{aligned}$\\
	\hline	
	
	%3.74
	$\!\begin{aligned}[c]
	& \texttt{$\vdash$ $\forall$ A B. }\\
	&\texttt{D\_OR (D\_OR (D\_BEFORE A B)}\\
	&\texttt {(D\_SIMULT A B))(D\_BEFORE B A) = D\_OR A B}\end{aligned}$& 
	$\!\begin{aligned}[b]
	(A \lhd B)+(A\Delta B)+(B \lhd A) \\= A+B
	\end{aligned}$\\
	\hline

%3.75
	$\!\begin{aligned}[c]
	& \texttt{$\vdash$ $\forall$ A B. }\\
	&\texttt{D\_OR (D\_OR(D\_AND A (D\_BEFORE B A))}\\
	&\texttt {(D\_SIMULT A B))(D\_AND B (D\_BEFORE A B)}\\
	&\texttt{= D\_AND A B}\end{aligned}$& 
	$\!\begin{aligned}[b]
	(A.(B\lhd A))+(A \Delta B)+\\(B.(A\lhd B)) = A.B
	\end{aligned}$\\
	\hline	
%3.76
	$\!\begin{aligned}[c]
	& \texttt{$\vdash$ $\forall$ A B. }\\
	&\texttt{D\_OR (D\_OR (D\_BEFORE A B)(D\_SIMULT A B))}\\
	&\texttt {(D\_AND A (D\_BEFORE B A)) = A}\end{aligned}$& 
	$\!\begin{aligned}[b]
	(A\lhd B)+(A \Delta B)\\+(A.(B \lhd A))= A
	\end{aligned}$\\
	\hline		
	
%3.77
	$\!\begin{aligned}[c]
	& \texttt{$\vdash$ $\forall$ A B C. }\\
	&\texttt{D\_AND (D\_AND (D\_BEFORE A B)}\\
	&\texttt {(D\_BEFORE B C))(D\_INCLUSIVE\_BEFORE A C)}\\
	&\texttt{= D\_AND (D\_BEFORE A B)(D\_BEFORE B C)}\end{aligned}$& 
	$\!\begin{aligned}[b]
	(A\lhd B).(B \lhd C).(A \unlhd C)=\\(A \lhd B).(B\lhd C)
	\end{aligned}$\\
	\hline		
\end{longtable}

\section{Experimental Results }
\label{Experiment results}
\begin{figure}[]
\centering
%\vspace{-20pt}
{\includegraphics[width=\linewidth , height=8cm]{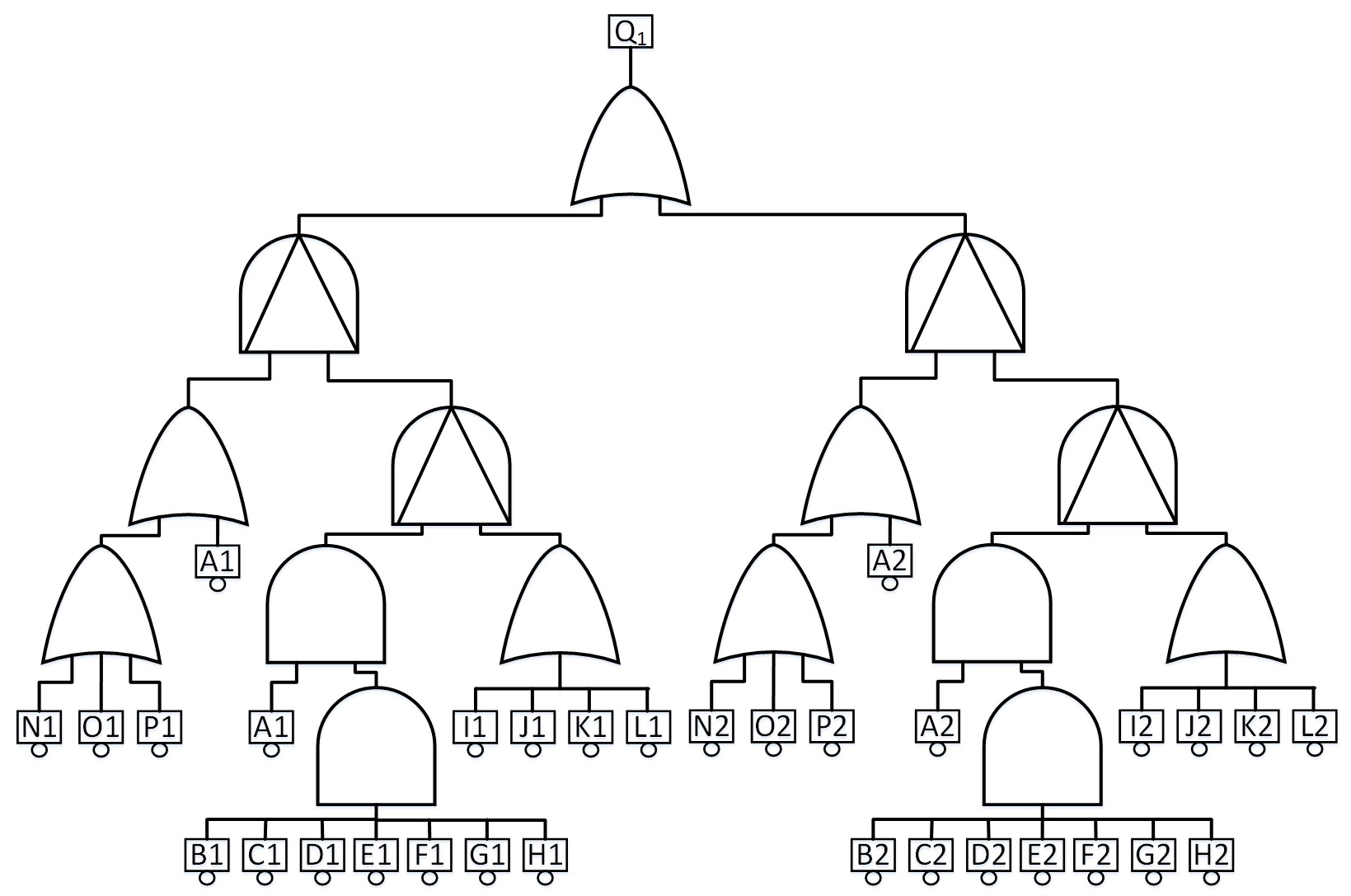}}
\caption{Scaled Cascaded PAND DFT}
%\vspace{-15pt}
\label{fig:Cascaded_PAND}
\end{figure}

In order to illustrate the effectiveness of the proposed methodology, we utilize it to conduct the formal DFT analysis of five benchmarks. 
The first benchmark, depicted in Figure \ref{fig:Cascaded_PAND}, is a scaled version of the original cascaded PAND fault tree \cite{boudali2007compositional , MerlePAND} with repeated events. In this work, we consider a scaled version of this DFT, i.e., two similar DFTs with different basic events and a top event that fails whenever one of these DFTs fails.

The second DFT is a modified and abstracted version of the Active Heat Rejection System (AHRS) \cite{Bayesian-Dugan}, which consists of two thermal rejection units $A$ and $B$. The failure of any of these two units leads to the failure of the whole system. Each main input ($A_{1}$ or $B_{1}$) has two spare parts, and the unit will fail with the failure of the main input and the spare inputs. All the inputs are functionally dependent on the power supply. 

\begin{figure}[]
\centering
%\vspace{-20pt}
{\includegraphics[width=0.7\linewidth , height=4cm]{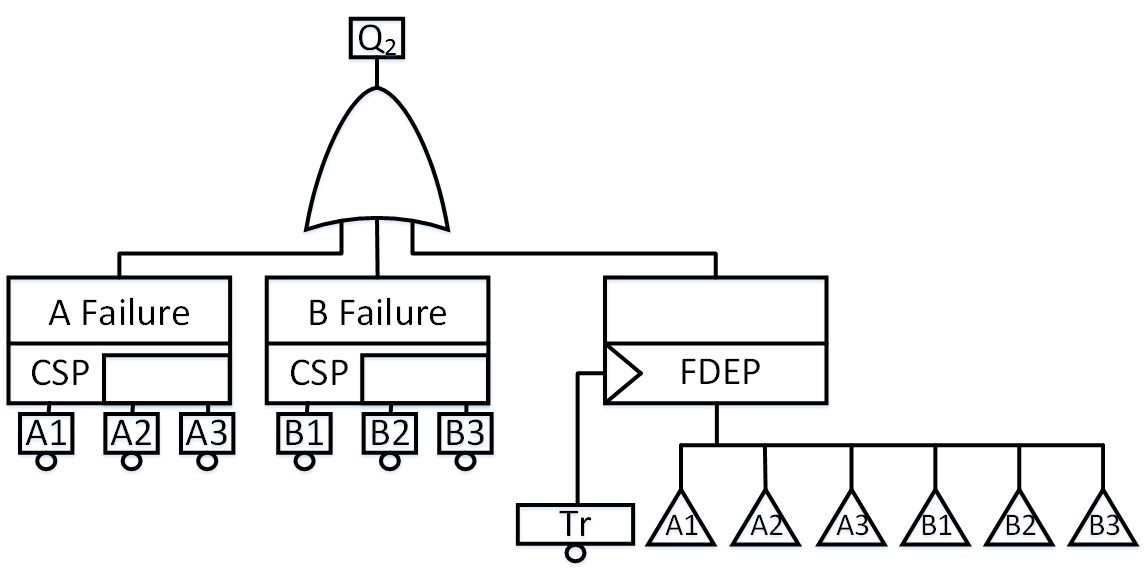}}
\caption{AHRS DFT}
%\vspace{-15pt}
\label{fig:AHRS}
\end{figure}
The third benchmark represents a Multiprocessor Computer System (MCS) \cite{TrivediMCS1995 , boudali2007compositional} with two redundant computers, having a processor, a disk and a memory unit. Each disk has its own spare and the two memory units share the same spare. The two processors are functionally dependent on the power supply. 

\begin{figure}[]
\centering
%\vspace{-20pt}
{\includegraphics[width=\linewidth , height=8cm]{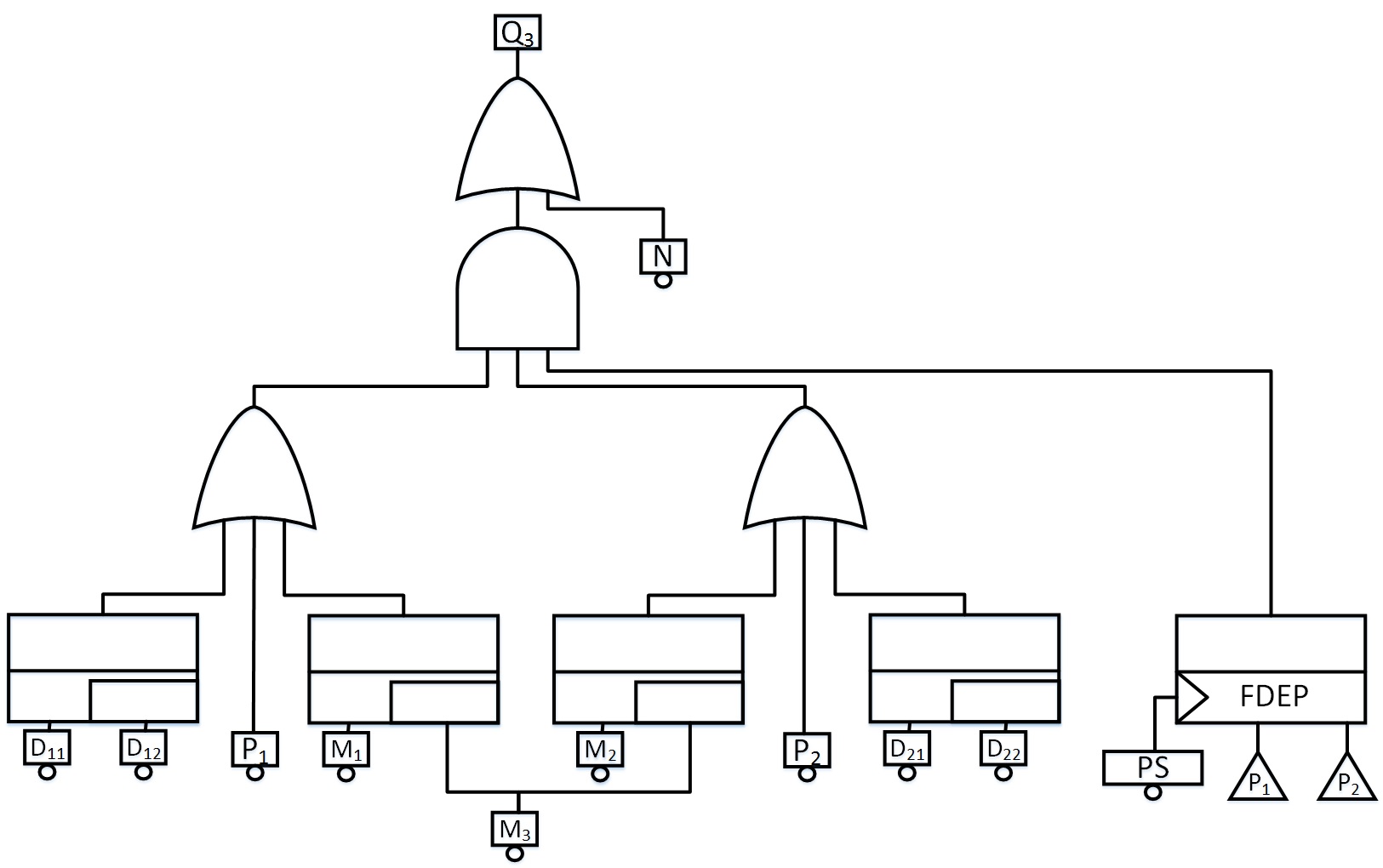}}
\caption{MCS DFT}
%\vspace{-15pt}
\label{fig:MCS}
\end{figure}
The fourth benchmark is a Hypotheical Example Computer System (HECS) \cite{DFT-handbook} consisting of two processors with a cold spare, five memory units, which are functionally dependent on two memory interface units and two system buses. The failure of the system also depends on the application subsystem, which in turn depends on the software, the hardware and the human operator. 

\begin{figure}[]
\centering
\vspace{-20pt}
{\includegraphics[width=\linewidth , height=9cm]{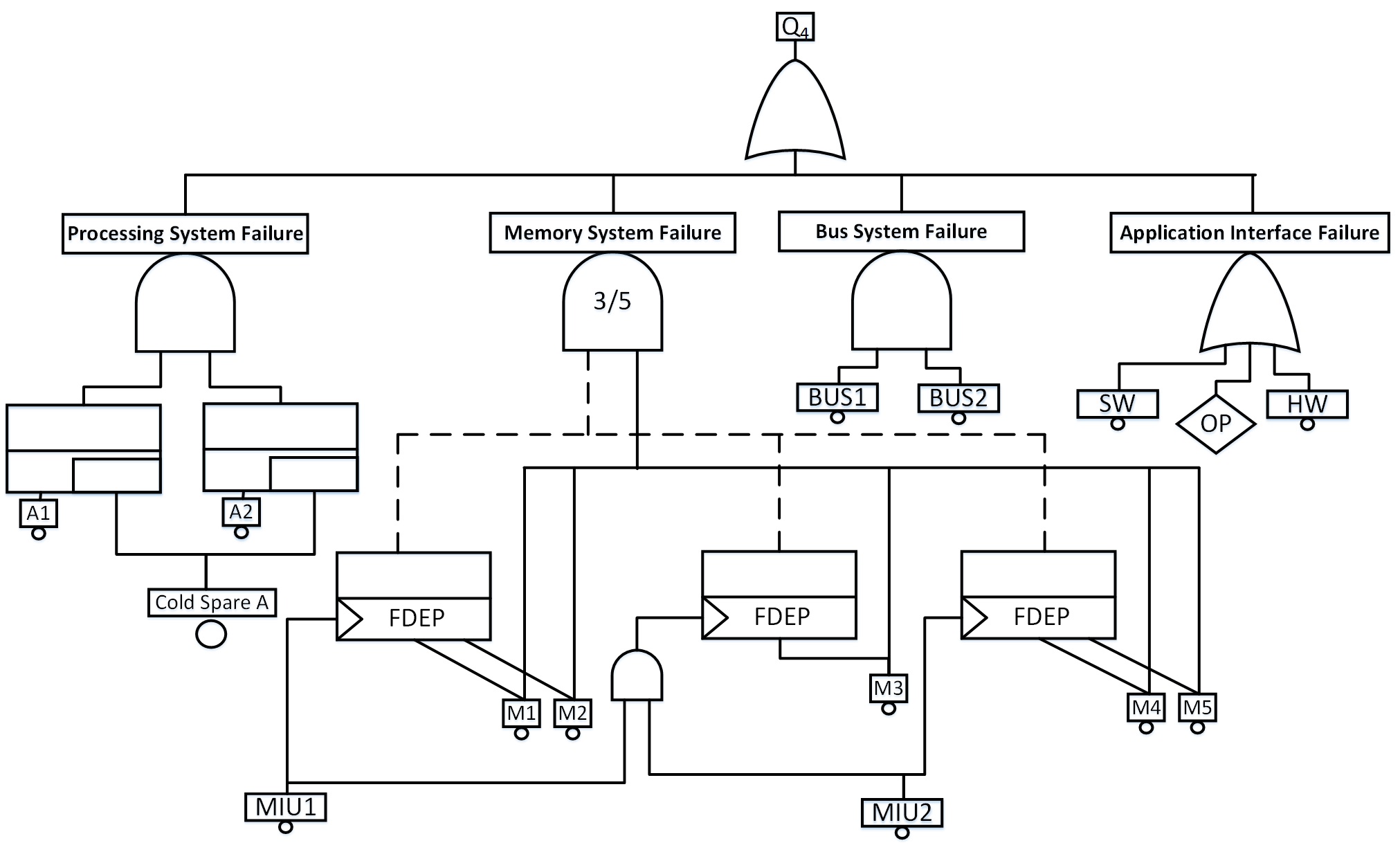}}
\caption{HECS DFT}
%\vspace{-15pt}
\label{fig:HECS}
\end{figure}
The last benchmark is a Hypothetical Cardiac Assist System (HCAS)\cite{boudali2007compositional , Merle-thesis}, which consists of two bumps ($P_{1}$ and $P_{2}$) with a shared spare ($BP$), two motors and a CPU ($P$) with a spare ($B$). Both CPUs are functionally dependent on a trigger, which represents the crossbar switch (CS) and the system supervisor (SS). 

\begin{figure}[]
\centering
%\vspace{-20pt}
{\includegraphics[width=\linewidth , height=8cm]{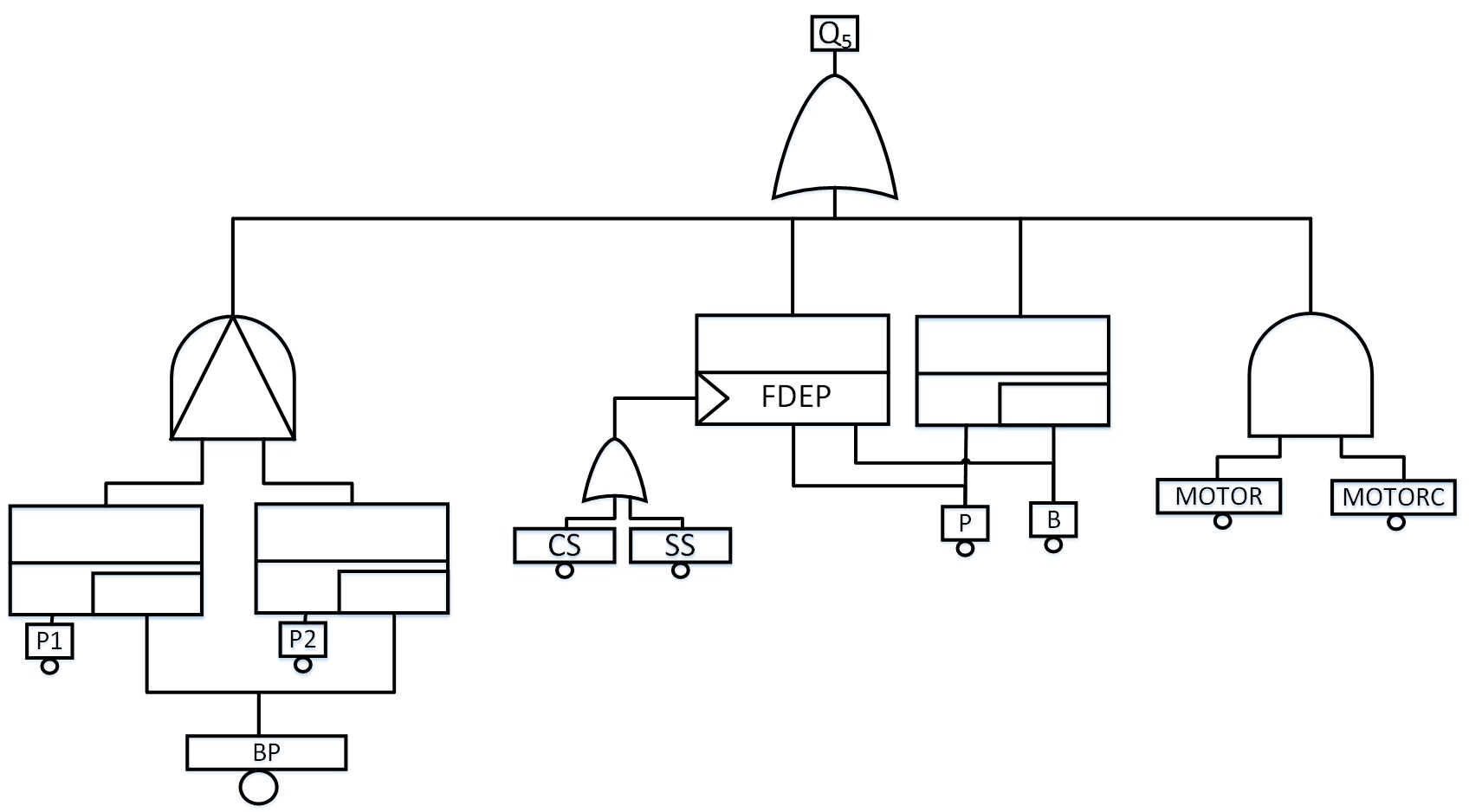}}
\caption{HCAS DFT}
%\vspace{-15pt}
\label{fig:HCAS}
\end{figure}
 
In the next section, the verification of the reduction will be introduced for each benchmark along with the reduced cut sets and sequences, then the quantitative analysis for the five benchmarks will be described in the subsequent section.

\subsection{Verifying the Reduced DFTs}
 
The first step in the proposed methodology is to create a formal model for both the original DFT and the reduced one. Then, these DFTs are checked if they are equal or not. After the equivalence verification, the cut sets and sequences can be determined. 

 \subsubsection{Verifying the Reduced Cascaded PAND DFT (CPAND) }

The top event ($Q_{1}$) of the system, shown in Figure \ref{fig:Cascaded_PAND}, is reduced using the simplification theorems as follows:
 \vspace{-4pt}
\begin{equation}
\label{eq:Cascaded_PAND_DFT_reduced1}
\begin{split}
Q_{1} = (I_{1} + J_{1} +K_{1} +L_{1} ).(A_{1} \lhd (I_{1} + J_{1} + K_{1} + L_{1} ). \\
((B_{1} C_{1} . D_{1} . E_{1} .W_{1} .G_{1} .H_{1})  \lhd ((I_{1} + J_{1} +K_{1} +L_{1}) \\
+ (I_{2} + J_{2} +K_{2} +L_{2} ).(A_{2} \lhd (I_{2} + J_{2} + K_{2} + L_{2} ). \\
((B_{2} C_{2} . D_{2} . E_{2} .W_{2} .G_{2} .H_{2})  \lhd ((I_{2} + J_{2} +K_{2} +L_{2})
\end{split}
\end{equation}
\vspace{-4pt}

\noindent We verify this simplification in HOL as follows:
\begin{thm}
\label{Cascaded_PAND_DFT_reduced}
\vspace{1pt} \texttt{$\vdash$ $\forall$ A1 B1 C1 D1 E1 W1 G1 H1 I1 J1 K1 L1 N1 O1 P1 A2 B2 C2 D2 E2 W2 G2 H2 I2 J2 K2 L2 N2 O2 P2. \\ALL\_DISTINCT [A1;B1;C1;D1;E1;W1;G1;H1;I1;J1;K1;L1;N1;O1;P1;A2;B2;C2;D2;E2;W2;G2;H2;I2;\\J2;K2;L2;N2;O2;P2] $\Rightarrow$\\
(D\_OR (PAND (D\_OR (A1)(D\_OR(D\_OR N1 O1) P1))(PAND (D\_AND A1 (D\_AND (D\_AND (D\_AND (D\_AND (D\_AND (D\_AND B1 C1) D1) E1) W1) G1) H1))(D\_OR (D\_OR (D\_OR I1 J1) K1) L1)))(PAND (D\_OR  A2 (D\_OR(D\_OR N2 O2) P2))(PAND (D\_AND A2 (D\_AND (D\_AND (D\_AND (D\_AND (D\_AND \\(D\_AND B2 C2) D2) E2) W2) G2) H2))
(D\_OR (D\_OR (D\_OR I2 J2) K2) L2)))=\\ D\_OR (D\_AND (D\_OR (D\_OR (D\_OR I1 J1) K1) L1)(D\_AND (D\_BEFORE A1 (D\_OR (D\_OR (D\_OR I1 J1) K1) L1))
(D\_BEFORE (D\_AND (D\_AND (D\_AND (D\_AND (D\_AND (D\_AND B1 C1) D1)E1) W1) G1) H1) (D\_OR (D\_OR (D\_OR I1 J1) K1) L1))))(D\_AND (D\_OR (D\_OR (D\_OR I2 J2 ) K2 ) L2)
(D\_AND \\(D\_BEFORE A2 (D\_OR (D\_OR (D\_OR I2 J2) K2) L2))(D\_BEFORE (D\_AND (D\_AND (D\_AND (D\_AND \\(D\_AND (D\_AND  B2 C2 ) D2) E2) W2) G2) H2) (D\_OR (D\_OR (D\_OR I2 J2) K2) L2)))))}
\end{thm}
\vspace{4pt}

\noindent The predicate \texttt{ALL\_DISTINCT} ensures that the basic events cannot occur at the same time. This condition was found to be a prerequisite for the above-mentioned consequence.  
From this reduction, it can be noticed that the basic events ($N_{1}$, $O_{1}$, $P_{1}$, $N_{2}$, $O_{2}$, $P_{2}$) have no effect on the failure of the top event since they are eliminated in the reduction. Considering the cut sets and sequences, the top event can fail in two cases. The first case corresponds to the first product in the structure function, which implies that the output event occurs if any one of the basic events ($I_{1}$, $J_{1}$, $K_{1}$, $L_{1}$) occurs and $A_{1}$ occurs before all of them and the inputs ($B_{1}$, $C_{1}$, $D_{1}$, $E_{1}$, $W_{1}$, $G_{1}$, $H_{1}$) occur before the inputs ($I_{1}$, $J_{1}$, $K_{1}$, $L_{1}$). The second case represents the second product of the second subtree, which is similar to the first product but with different basic events. 
Since the Galileo format (which is used to model a DFT in STORM) supports only DFT gates and not operators, it is required that the reduced form is represented using DFT gates only. This representation is verified in HOL as follows:

\begin{thm}
\label{Cascaded_PAND_DFT_reduced}
\vspace{1pt} \texttt{$\vdash$ $\forall$ A1 B1 C1 D1 E1 W1 G1 H1 I1 J1 K1 L1 N1 O1 P1 A2 B2 C2 D2 E2 W2 G2 H2 I2 J2 K2 L2 N2 O2 P2. \\ALL\_DISTINCT [A1;B1;C1;D1;E1;W1;G1;H1;I1;J1;K1;L1;N1;O1;P1;A2;B2;C2;D2;E2;W2;G2;H2;I2;\\J2;K2;L2;N2;O2;P2] $\Rightarrow$\\
(D\_OR (PAND (D\_OR (A1)(D\_OR(D\_OR N1 O1) P1))(PAND (D\_AND A1 (D\_AND (D\_AND (D\_AND (D\_AND (D\_AND (D\_AND B1 C1) D1) E1) W1) G1) H1))(D\_OR (D\_OR (D\_OR I1 J1) K1) L1)))(PAND (D\_OR  A2 (D\_OR(D\_OR N2 O2) P2))(PAND (D\_AND A2 (D\_AND (D\_AND (D\_AND (D\_AND (D\_AND \\(D\_AND B2 C2) D2) E2) W2) G2) H2))
(D\_OR (D\_OR (D\_OR I2 J2) K2) L2)))=\\ 
D\_OR (D\_AND (PAND A1 (D\_OR (D\_OR (D\_OR I1 J1) K1) L1))(PAND (D\_AND (D\_AND (D\_AND (D\_AND (D\_AND (D\_AND B1 C1) D1)E1) W1) G1) H1)(D\_OR (D\_OR (D\_OR I1 J1) K1) L1)))
(D\_AND (PAND A2 (D\_OR (D\_OR (D\_OR I2 J2 ) K2 ) L2)(PAND (D\_AND (D\_AND (D\_AND (D\_AND (D\_AND (D\_AND  B2 C2 ) D2) E2) W2) G2) H2)(D\_OR (D\_OR (D\_OR I2 J2) K2) L2)))}
\end{thm}
\vspace{4pt}

\subsubsection{Verifying the Reduced AHRS DFT}

The top event ($Q_{2}$) of the system shown in Figure \ref{fig:AHRS} is reduced using the algebraic simplification theorems, assuming that the spares are cold spares:

\begin{equation}
\label{eq:AHRS}
\begin{split}
Q_{2} = Tr + A_{3a}.(A_{1} \lhd A_{2a}).(A_{2a} \lhd A_{3a}) + B_{3a} .(B_{1} \lhd B_{2a}).(B_{2a} \lhd  B_{3a})
\end{split}
\end{equation}

\noindent We verify this in HOL as:

\begin{thm}
\label{ARHS}
 \vspace{1pt} \texttt{$\vdash$ $\forall$ A1 A2\_a A2\_d A3\_a A3\_d Tr B1 B2\_a B2\_d B3\_a B3\_d.\\
       ALL\_DISTINCT [A1; A2\_a; A2\_d; A3\_a; A3\_d; Tr; B1; B2\_a; B2\_d; B3\_a; B3\_d] $\wedge$\\
       COLD\_SPARE A2\_d $\wedge$ COLD\_SPARE A3\_d $\wedge$ COLD\_SPARE B2\_d $\wedge$\  COLD\_SPARE B3\_d $\Rightarrow$\\
       (D\_OR (WSP (FDEP Tr A1)(WSP (FDEP A2\_a Tr) (FDEP A3\_a Tr) (FDEP Tr A3\_d))
             \\(WSP (FDEP Tr A2\_d) (FDEP A3\_a Tr) (FDEP A3\_d Tr)))
          (WSP (FDEP Tr A1)\\(WSP (FDEP A2\_a Tr)(FDEP A3\_a Tr) (FDEP Tr A3\_d))
             (WSP(FDEP Tr A2\_d) (FDEP A3\_a Tr)
               \\ (FDEP A3\_d Tr))) =\\
        D\_OR Tr (D\_OR (D\_AND (D\_AND A3\_a (D\_BEFORE A1 A2\_a))(D\_BEFORE A2\_a A3\_a))\\(D\_AND (D\_AND B3\_a (D\_BEFORE B1 B2\_a))(D\_BEFORE B2\_a B3\_a))))}
\end{thm}

This system has three sources of failure; the trigger, the sequence of failure of ($A_{1}$ then $A_{2}$ then $A_{3}$) and finally the sequence ($B_{1}$ then $B_{2}$ then $B_{3}$)

\subsubsection{Verifying the Reduced MCS DFT}

The top event ($Q_{3}$) of the system shown in Figure \ref{fig:MCS} is reduced using the algebraic simplification theorems to:

\begin{equation}
\label{eq:MCS}
\begin{split}
Q_{3} = N + PS +(P_{1} + D_{11}.(D_{12d}\lhd D_{11})+D_{12a}.(D_{11}\lhd D_{12a}\\
+ M_{1}.(M_{3d} \lhd M_{1})+M_{3a}.(M_{1} \lhd M_{3a})+M_{1}.(M_{2} \lhd M_{1} )\\
.(P_{2} + D_{21}.(D_{22d}\lhd D_{21})+D_{22a}.(D_{21}\lhd D_{22a}+ M_{2}.\\
(M_{3d} \lhd M_{2})+M_{3a}.(M_{2} \lhd M_{3a})+M_{2}.(M_{1} \lhd M_{2})
\end{split}
\end{equation}
\noindent We verify this in HOL as:

\begin{thm}
\label{MCS}
 \vspace{1pt} \texttt{$\vdash$ $\forall$ M1 M2 M3\_a M3\_d D11 D12\_a D12\_d D21 D22\_a D22\_d P1 P2 PS N.\\
       ALL\_DISTINCT [M1; M2; M3\_a; M3\_d; D11; D12\_a; D12\_d; D21; D22\_a; D22\_d; P1; P2; PS; N] $\Rightarrow$\\
       (D\_OR N (D\_AND (D\_OR (D\_OR (WSP D11 D12\_a D12\_d) (FDEP PS P1))
                \\(shared\_spare M1 M2 M3\_a M3\_d))
             (D\_OR (D\_OR (WSP D21 D22\_a D22\_d)(FDEP PS P2))
                \\(shared\_spare M2 M1 M3\_a M3\_d))) =\\
        D\_OR (D\_OR N PS)(D\_AND
             (D\_OR (D\_OR P1 (WSP D11 D12\_a D12\_d))
                \\(shared\_spare M1 M2 M3\_a M3\_d))
             (D\_OR (D\_OR P2 (WSP D21 D22\_a D22\_d))
                \\(shared\_spare M2 M1 M3\_a M3\_d))))
}
\end{thm}

From this verified reduced function, the sources of failure are: $N$, $PS$ or the failure of both computers by the failure of any element in each one.

\subsubsection{Verifying the Reduced HECS DFT}

The top event ($Q_{4}$) of the system shown in Figure \ref{fig:HECS} is reduced using the algebraic simplification theorems to\cite{Merle-thesis}:

\begin{equation}
\label{eq:HECS}
\begin{split}
Q_{4} = A_{a}.A_{2}.(A_{1} \lhd A_{a}).(A_{1} \lhd A_{2})+ A_{a}.A_{1}.(A_{2} \lhd A_{a}).(A_{2} \lhd A_{1})\\+ MIU_{1}.MIU_{2}+MIU_{1}.M_{3}+MIU_{1}.M_{4}+MIU_{1}.M_{5}\\+MIU_{2}.M_{1}+MIU_{2}.M_{2}+MIU_{2}.M_{3}+M_{1}.M_{2}.M_{3}\\+M_{1}.M_{2}.M_{4}+M_{1}.M_{2}.M_{5}+M_{1}.M_{3}.M_{4}+M_{1}.M_{3}.M_{5}\\+M_{1}.M_{4}.M_{5}+M_{2}.M_{3}.M_{4}+M_{2}.M_{3}.M_{5}+\\M_{2}.M_{4}.M_{5}+M_{3}.M_{4}.M_{5}+BUS_{1}.BUS_{2}+HW+SW+OP
\end{split}
\end{equation}
\noindent We verify this in HOL as:

\begin{thm}
\label{HECS}
\vspace{1pt} \texttt{$\vdash$ $\forall$ A1 A2 A\_a A\_d M1 M2 M3 M4 M5 MIU1 \\MIU2 OP HW SW BUS1 BUS2.
       NEVER\_events A\_a A\_d $\wedge$ COLD\_SPARE A\_d $\wedge$\\
       ALL\_DISTINCT[A1; A2 ;A\_a; A\_d; M1 ;M2; M3; M4; M5; MIU1; MIU2; OP; HW; SW; BUS1; \\BUS2]$\Rightarrow$\\
       (D\_OR
          (D\_OR
             (D\_OR
                (D\_AND (shared\_spare A1 A2 A\_a A\_d)
                   (shared\_spare A2 A1 A\_a A\_d))
                \\(D3of5 (FDEP M5 MIU2) (FDEP M4 MIU2)
                   (FDEP M3 (D\_AND MIU2 MIU1)) (FDEP M2 MIU1)
                  \\ (FDEP M1 MIU1))) (D\_AND BUS1 BUS2))
          (D\_OR (D\_OR SW HW) OP) =\\
        D\_OR
          (D\_OR
             (D\_OR
                (D\_OR
                   (D\_AND (D\_AND (D\_AND A\_a A2) (D\_BEFORE A1 A\_a))
                     \\ (D\_BEFORE A1 A2))
                   (D\_AND (D\_AND (D\_AND A\_a A1) (D\_BEFORE A2 A\_a))
                      (D\_BEFORE A2 A1)))\\
                (D\_OR
                   (D\_OR
                      (D\_OR
                         (D\_OR
                            (D\_OR
                               (D\_OR
                                  (D\_OR
                                     (D\_OR
                                        (D\_OR
                                           (D\_OR
                                              (D\_OR
                                                 (D\_OR
                                                    (D\_OR
                                                       (D\_OR\\
                                                          (D\_OR
                                                             (D\_OR
                                                                (D\_AND
                                                                   MIU1
                                                                   MIU2)
                                                                (D\_AND
                                                                   MIU1
                                                                   M3))
                                                             (D\_AND MIU1
                                                                M4))
                                                          (D\_AND MIU1
                                                             M5))\\
                                                       (D\_AND MIU2 M1))
                                                    (D\_AND MIU2 M2))
                                                 (D\_AND MIU2 M3))
                                              (D\_AND (D\_AND M1 M2) M3))\\
                                           (D\_AND (D\_AND M1 M2) M4))
                                        (D\_AND (D\_AND M1 M2) M5))
                                     (D\_AND (D\_AND M1 M3) M4))\\
                                  (D\_AND (D\_AND M1 M3) M5))
                               (D\_AND (D\_AND M1 M4) M5))
                            (D\_AND (D\_AND M2 M3) M4))\\
                         (D\_AND (D\_AND M2 M3) M5))
                      (D\_AND (D\_AND M2 M4) M5))
                   (D\_AND (D\_AND M3 M4) M5)))\\ (D\_AND BUS1 BUS2))
          (D\_OR (D\_OR SW HW) OP))}
\end{thm}
\noindent where \texttt{D3of5} is a combination of AND and OR operators to create the voting 3 out of 5 gate. The cut sets and sequences of this function can be easily obtained from Equation \ref{eq:HECS}. To be able to model this function in STORM, it was verified in HOL in terms of the dynamic gates as:

\begin{thm}
\label{HECS1}
\vspace{1pt} \texttt{$\vdash$ $\forall$ A1 A2 A\_a A\_d M1 M2 M3 M4 M5 MIU1 MIU2 OP HW SW BUS1 BUS2.\\
       NEVER\_events A\_a A\_d $\wedge$ COLD\_SPARE A\_d $\wedge$\\
       ALL\_DISTINCT[A1; A2 ;A\_a; A\_d; M1 ;M2; M3; M4; M5; MIU1; MIU2; OP; HW; SW; BUS1; \\BUS2]$\Rightarrow$\\
       (D\_OR
          (D\_OR
             (D\_OR
                (D\_AND (shared\_spare A1 A2 A\_a A\_d)
                   (shared\_spare A2 A1 A\_a A\_d))
                \\(D3of5 (FDEP M5 MIU2) (FDEP M4 MIU2)
                   (FDEP M3 (D\_AND MIU2 MIU1)) (FDEP M2 MIU1)
                  \\ (FDEP M1 MIU1))) (D\_AND BUS1 BUS2))
          (D\_OR (D\_OR SW HW) OP) =\\
        D\_OR
          (D\_OR
             (D\_OR
                (D\_OR
                (D\_AND (shared\_spare A1 A2 A\_a A\_d)
                   \\(shared\_spare A2 A1 A\_a A\_d))
                (D\_OR
                   (D\_OR
                      (D\_OR
                         (D\_OR
                            (D\_OR
                               (D\_OR
                                  (D\_OR
                                     (D\_OR
                                        (D\_OR\\
                                           (D\_OR
                                              (D\_OR
                                                 (D\_OR
                                                    (D\_OR
                                                       (D\_OR
                                                          (D\_OR
                                                             (D\_OR
                                                                (D\_AND
                                                                   MIU1
                                                                   MIU2)
                                                                (D\_AND
                                                                   MIU1
                                                                   M3))\\
                                                             (D\_AND MIU1
                                                                M4))
                                                          (D\_AND MIU1
                                                             M5))
                                                       (D\_AND MIU2 M1))
                                                    (D\_AND MIU2 M2))
                                                 (D\_AND MIU2 M3))\\
                                              (D\_AND (D\_AND M1 M2) M3))
                                           (D\_AND (D\_AND M1 M2) M4))
                                        (D\_AND (D\_AND M1 M2) M5))\\
                                     (D\_AND (D\_AND M1 M3) M4))
                                  (D\_AND (D\_AND M1 M3) M5))
                               (D\_AND (D\_AND M1 M4) M5))\\
                            (D\_AND (D\_AND M2 M3) M4))
                         (D\_AND (D\_AND M2 M3) M5))
                      (D\_AND (D\_AND M2 M4) M5))\\
                   (D\_AND (D\_AND M3 M4) M5))) (D\_AND BUS1 BUS2))
          (D\_OR (D\_OR SW HW) OP))}
\end{thm}

\subsubsection{Verifying the Reduced HCAS DFT}

The top event ($Q_{5}$) of the system shown in Figure \ref{fig:HCAS} is reduced using the algebraic simplification theorems to\cite{Merle-thesis}:

\begin{equation}
\label{eq:HCAS}
\begin{split}
Q_{5} = CS+SS+MOTOR.MOTORC+P.(B_{d} \lhd P)+B_{a}.(P \lhd B_{a}) \\+BP_{a}.(P_{2} \lhd P_{1}).(P_{1} \lhd BP_{a})+P_{2}.(P_{1} \lhd BP_{a}).(BP_{a} \lhd P_{2})
\end{split}
\end{equation}
\noindent We verify this in HOL as:
\begin{thm}
\label{HCAS}
\vspace{1pt}\texttt{$\vdash$ $\forall$ CS SS MOTOR MOTORC P B\_d B\_a BP\_a BP\_d P1 P2. NEVER\_events B\_a B\_d $\wedge$\\
       ALL\_DISTINCT [P1; P2; BP\_a; BP\_d; P; B\_a; B\_d; CS; SS; MOTOR; MOTORC]$\wedge$ \\(D\_BEFORE B\_a P = NEVER) $\wedge$
       (D\_AND (D\_BEFORE BP\_a P1) (D\_BEFORE P1 P2) = NEVER) $\wedge$\\
       (D\_AND (D\_BEFORE BP\_a P2) (D\_BEFORE P2 P1) = NEVER) $\wedge$\\
       NEVER\_events BP\_a BP\_d $\wedge$ COLD\_SPARE BP\_d $\Rightarrow$\\
       (D\_OR
          (D\_OR
             (WSP (FDEP P (D\_OR CS SS)) (FDEP B\_a (D\_OR CS SS))
               \\ (FDEP B\_d (D\_OR CS SS))) (D\_AND MOTOR MOTORC))
          (PAND (shared\_spare P1 P2 BP\_a BP\_d)
            \\(shared\_spare P2 P1 BP\_a BP\_d)) =\\
        D\_OR
          (D\_OR
             (D\_OR
                (D\_OR (D\_OR (D\_OR CS SS) (D\_AND MOTOR MOTORC))
                  \\ (D\_AND P (D\_BEFORE B\_d P)))
                (D\_AND B\_a (D\_BEFORE P B\_a)))
             (D\_AND (D\_AND BP\_a \\(D\_BEFORE P2 P1)) (D\_BEFORE P1 BP\_a)))
          (D\_AND (D\_AND P2 (D\_BEFORE P1 BP\_a)) \\(D\_BEFORE BP\_a P2)))
          }
          \end{thm}

The cut sets and sequences for $Q_{5}$ can be obtained from the verified reduced function. To model this function in STORM, it was verified using the dynamic gates, assuming that $B$ is a cold spare:
\begin{thm}
\label{HCAS}
\vspace{1pt}\texttt{$\vdash$ $\forall$ CS SS MOTOR MOTORC P B\_d B\_a BP\_a BP\_d P1 P2. NEVER\_events B\_a B\_d $\wedge$\\
       ALL\_DISTINCT [P1; P2; BP\_a; BP\_d; P; B\_a; B\_d; CS; SS; MOTOR; MOTORC]$\wedge$ \\(D\_BEFORE B\_a P = NEVER) $\wedge$
       (D\_AND (D\_BEFORE BP\_a P1) (D\_BEFORE P1 P2) = NEVER) $\wedge$\\
       (D\_AND (D\_BEFORE BP\_a P2) (D\_BEFORE P2 P1) = NEVER) $\wedge$\\
       NEVER\_events BP\_a BP\_d $\wedge$ COLD\_SPARE BP\_d $\Rightarrow$\\
       (D\_OR
          (D\_OR
             (WSP (FDEP P (D\_OR CS SS)) (FDEP B\_a (D\_OR CS SS))
               \\ (FDEP B\_d (D\_OR CS SS))) (D\_AND MOTOR MOTORC))
          (PAND (shared\_spare P1 P2 BP\_a BP\_d)
            \\(shared\_spare P2 P1 BP\_a BP\_d)) =\\
        D\_OR
          (D\_OR
             (D\_OR
                 (D\_OR CS SS) (D\_AND MOTOR MOTORC))
                    (D\_AND B\_a (D\_BEFORE P B\_a)))
             \\(PAND (shared\_spare P1 P2 BP\_a BP\_d)
             (shared\_spare P2 P1 BP\_a BP\_d)))
          }
          \end{thm}

 \subsection{Probabilistic Analysis Results using STORM}
 \vspace{-4pt}
The quantitative analysis for the five benchmarks was conducted using STORM on a Linux machine with i7 2.4 GHZ quad core CPU and 4 GB of RAM. The efficiency of the proposed methodology is highlighted by analyzing the original DFTs and the reduced ones. In addition, the probability of failure for each DFT is evaluated for different time bounds, e.g.  the probability of failure after 100 working time units. A summary of the analysis results are given in Table \ref{table:storm_results}. It can be noticed that the number of states is reduced as well as the total analysis time. For the first benchmark, the analysis time is reduced due to the huge reduction in the number of states. As mentioned earlier, many basic events are eliminated using the algebraic reduction theorems, which in turn reduced the total analysis time as well as the number of states. For the rest of the benchmarks, the analysis time is significantly reduced when the reduced DFT is used in the analysis. This is mainly because of two reasons, firstly, the number of states is reduced, and secondly, the original DFT is modeled as a Markov Automata (MA) as there is a non-deterministic behavior, while the reduced DFT is modeled as a Continuous Time Markov Chain (CTMC). This means that in the reduced DFT the non-deterministic behavior caused by the failure dependency does not exist any more, as the reduction process depends on the time of failure of the gates. We used the STORM command (firstdep)\cite{STORM} to resolve the non-deterministic behavior in the original DFT to generate a CTMC instead of a MA, and the results in Table \ref{table:storm_analysis} show that the number of states for the reduced DFTs is generally smaller than that of the original DFT with resolved dependencies, which emphasizes on the importance of the proposed methodology not only in providing a formal qualitative analysis but also in reducing the quantitative analysis cost in terms of time and memory, i.e., number of states.
\vspace{30pt}

\begin{table}[h]
\centering
\vspace{-5pt}
\caption{STORM Analysis Results (Before and After Reduction)}

\label{table:storm_results}

\resizebox{\textwidth}{!}{%
\begin{tabular}{@{}lllllllll@{}}
\cmidrule(r){1-9}% \cmidrule(l){7-9}
$DFT$ \phantom{ab}                & $Time$\phantom{ab} & \multicolumn{3}{c}{$Before\ Reduction$} &\phantom{abcd}   & \multicolumn{3}{c}{$After\ Reduction$} \\ \cmidrule(lr){3-5} \cmidrule(l){7-9} 
                  & $Bound$\phantom{ab} & $\#States$\phantom{abcdefijklm}      & $Analysis$ \phantom{ab}     &$ Probability$\phantom{ab}     &  & $\#States$\phantom{abcdefijk}      & $Analysis$ \phantom{ab}    & $Probability$ \phantom{ab}    \\
                  &   &          & $Time(sec)$\phantom{ab}      &   $of Failure$     &  &          & $Time(sec)$ \phantom{ab}    &  $of Failure$      \\ \cmidrule(r){1-9}% \cmidrule(l){7-9} 
%\multirow{3}{*}{CPAND} & \cellcolor[HTML]{EFEFEF}1000  &  \cellcolor[HTML]{EFEFEF}  148226 (CTMC)     &\cellcolor[HTML]{EFEFEF}   7.488     & \cellcolor[HTML]{EFEFEF}1.464103531e-4       &\cellcolor[HTML]{EFEFEF}  &  \cellcolor[HTML]{EFEFEF}66050 (CTMC)  & \cellcolor[HTML]{EFEFEF}3.032   &\cellcolor[HTML]{EFEFEF} 1.464103348e-4       \\
CPAND & 1000  &    148226 (CTMC)     &   7.488     & 1.464103531e-4       &  &  66050 (CTMC)  & 3.032   & 1.464103348e-4       \\

              %    & 10  &          &        &        &  &          &       &        \\
              %    &  100 &          &        &        &  &          &       &        \\
               \cmidrule(r){1-9} %\cmidrule(l){7-9} 
\multirow{2}{*}{ARHS} %& 1  &   74 (MA)     & 1.874 &  9.995004906e-4      &  &  10 (CTMC) & 0.067& 9.994968554e-4      \\

                  &  \cellcolor[HTML]{EFEFEF}10  &   \cellcolor[HTML]{EFEFEF}74 (MA)     & \cellcolor[HTML]{EFEFEF}169.81      & \cellcolor[HTML]{EFEFEF}0.00995049462      & \cellcolor[HTML]{EFEFEF} &\cellcolor[HTML]{EFEFEF}10 (CTMC)   & \cellcolor[HTML]{EFEFEF}0.067   &  \cellcolor[HTML]{EFEFEF}0.009950461197     \\

                  &  100 &   74 (MA)     & *       & **       &  &   10 (CTMC)      &  0.067     & 0.954423939       \\
 \cmidrule(r){1-9} %\cmidrule(l){7-9} 

\multirow{2}{*}{MCS} %&\cellcolor[HTML]{EFEFEF}1  &\cellcolor[HTML]{EFEFEF}89 (MA)     &\cellcolor[HTML]{EFEFEF}1.587 &\cellcolor[HTML]{EFEFEF}0.001199640348      &\cellcolor[HTML]{EFEFEF}  &\cellcolor[HTML]{EFEFEF}29 (CTMC) &\cellcolor[HTML]{EFEFEF}0.042&\cellcolor[HTML]{EFEFEF}0.001199627432      \\

                  & \cellcolor[HTML]{EFEFEF}10  &  \cellcolor[HTML]{EFEFEF}89 (MA)    \cellcolor[HTML]{EFEFEF} &  \cellcolor[HTML]{EFEFEF}139.7      & \cellcolor[HTML]{EFEFEF}0.01196434683      &\cellcolor[HTML]{EFEFEF}& \cellcolor[HTML]{EFEFEF}29 (CTMC)    & \cellcolor[HTML]{EFEFEF}0.061   &  \cellcolor[HTML]{EFEFEF}0.01196434516     \\

                  &100 &89 (MA)     &*       &**       &&29 (CTMC)      &0.060     &0.1166464887       \\
 \cmidrule(r){1-9} %\cmidrule(l){7-9} 
                    
\multirow{2}{*}{HECS} %& 1  &   1051 (MA)      & 32.176 &  0.001701070343      &  &  505 (CTMC) & 0.119& 0.001701011276      \\

                  &\cellcolor[HTML]{EFEFEF}10  &\cellcolor[HTML]{EFEFEF}1051 (MA)     &\cellcolor[HTML]{EFEFEF}16359.83      &\cellcolor[HTML]{EFEFEF}0.01710278909      &\cellcolor[HTML]{EFEFEF}&\cellcolor[HTML]{EFEFEF}505 (CMTC)   &\cellcolor[HTML]{EFEFEF}0.123   &\cellcolor[HTML]{EFEFEF}0.01710276373    \\

                  &  100 &   1051 (MA)     & *       & **       &  &   505 (CMTC)      &  0.123     & 0.1762782397       \\
 \cmidrule(r){1-9} %\cmidrule(l){7-9} 
        
\multirow{2}{*}{HCAS} %%&\cellcolor[HTML]{EFEFEF}1  &\cellcolor[HTML]{EFEFEF}181 (MA)     &\cellcolor[HTML]{EFEFEF}3.02 &\cellcolor[HTML]{EFEFEF}2.0000102e-6      &\cellcolor[HTML]{EFEFEF}&\cellcolor[HTML]{EFEFEF}37 (CTMC)  &\cellcolor[HTML]{EFEFEF}0.064&\cellcolor[HTML]{EFEFEF}2.9999929621e-6      \\

                  & \cellcolor[HTML]{EFEFEF}10  &    \cellcolor[HTML]{EFEFEF}181 (MA)    &  \cellcolor[HTML]{EFEFEF}275.31      &  \cellcolor[HTML]{EFEFEF}2.000104327e-5      &\cellcolor[HTML]{EFEFEF}  & \cellcolor[HTML]{EFEFEF}37 (CTMC)   & \cellcolor[HTML]{EFEFEF}0.070   & \cellcolor[HTML]{EFEFEF}2.99929683e-5     \\

                  &100 &181 (MA)    &*       &**       &&37 (CTMC)      &0.071     &0.000300083976       \\
 \cmidrule(r){1-9} %\cmidrule(l){7-9} 
                      
\end{tabular}%
}
\begin{tablenotes}
\scriptsize
\item [*] * The analysis did not finish within 4 hours 
\item [**]** No probabilities are recorded (analysis did not finish)
\vspace{-1em}
\end{tablenotes}
\end{table}
%\vspace{-35pt}
\begin{table}[h]
\centering
\caption{STORM Analysis Results with Resolved Dependencies}
\label{table:storm_analysis}
\resizebox{\textwidth}{!}{%
\begin{tabular}{@{}lllllllll@{}}
\cmidrule(r){1-9}% \cmidrule(l){7-9}
$DFT$ \phantom{ab}                & $Time$\phantom{ab} & \multicolumn{3}{c}{$Dependency\ resolved\ in\ STORM$} & \phantom{abcd}  & \multicolumn{3}{c}{$Algebraic\ Reduction$} \\ \cmidrule(lr){3-5} \cmidrule(l){7-9} 
                  & $Bound$\phantom{ab} & $\#States$\phantom{abcdefijklm}      & $Analysis$ \phantom{ab}     &$ Probability$\phantom{ab}     &  & $\#States$\phantom{abcdefijk}      & $Analysis$ \phantom{ab}    & $Probility$ \phantom{ab}    \\
                  &   &          & $Time(sec)$\phantom{ab}      & $of Failure$      &  &          & $Time(sec)$ \phantom{ab}    &   $of Failure$     \\ \cmidrule(r){1-9}% \cmidrule(l){7-9} 
%\cmidrule(l){7-9} 
\multirow{2}{*}{ARHS} % & 1  &   10 (CTMC)     & 0.069 &  9.99496855e-4      &  &  10 (CTMC) & 0.067& 9.994968554e-4      \\

                  &  \cellcolor[HTML]{EFEFEF}10  &   \cellcolor[HTML]{EFEFEF}10(CTMC)     & \cellcolor[HTML]{EFEFEF}0.068     & \cellcolor[HTML]{EFEFEF}0.009960461197     & \cellcolor[HTML]{EFEFEF} &\cellcolor[HTML]{EFEFEF}10 (CTMC)   & \cellcolor[HTML]{EFEFEF}0.067   &  \cellcolor[HTML]{EFEFEF}0.009950461197     \\

                  &  100 &   10 (CTMC)     & 0.1      & 0.09544239393       &  &   10 (CTMC)      &  0.067     & 0.954423939       \\
 \cmidrule(r){1-9} %\cmidrule(l){7-9} 

\multirow{2}{*}{MCS} %&\cellcolor[HTML]{EFEFEF}1  &\cellcolor[HTML]{EFEFEF}45 (CTMC)     &\cellcolor[HTML]{EFEFEF} 0.071 &\cellcolor[HTML]{EFEFEF} 0.001199627423      &\cellcolor[HTML]{EFEFEF}  &\cellcolor[HTML]{EFEFEF}29 (CTMC) &\cellcolor[HTML]{EFEFEF}0.042&\cellcolor[HTML]{EFEFEF}0.001199627432      \\

                  & \cellcolor[HTML]{EFEFEF}10  &    \cellcolor[HTML]{EFEFEF}45 (CMTC)     & \cellcolor[HTML]{EFEFEF}0.064      & \cellcolor[HTML]{EFEFEF}0.01196434516      &\cellcolor[HTML]{EFEFEF}  &\cellcolor[HTML]{EFEFEF}29 (CTMC)    &\cellcolor[HTML]{EFEFEF}0.061   &  \cellcolor[HTML]{EFEFEF}0.01196434516     \\

                  &100 &45(CMTC)     &0.064     &0.1166464887       &&29 (CTMC)      &0.060     &0.1166464887       \\
 \cmidrule(r){1-9} %\cmidrule(l){7-9} 
                    
\multirow{2}{*}{HECS} % & 1  &   379(CTMC)      & 0.117 &  0.001701011276      &  &  505 (CTMC) & 0.119& 0.001701011276      \\

                  &\cellcolor[HTML]{EFEFEF}10  &\cellcolor[HTML]{EFEFEF}379 (CTMC)     &\cellcolor[HTML]{EFEFEF}0.118    &\cellcolor[HTML]{EFEFEF}0.01710276373     &\cellcolor[HTML]{EFEFEF}&\cellcolor[HTML]{EFEFEF}505 (CMTC)   &\cellcolor[HTML]{EFEFEF}0.123   &\cellcolor[HTML]{EFEFEF}0.01710276373    \\

                  &  100 &  379 (CTMC)     & 0.121       & 0.1762782397      &  &   505 (CMTC)      &  0.123     & 0.1762782397       \\
 \cmidrule(r){1-9} %\cmidrule(l){7-9} 
        
\multirow{4}{*}{HCAS} %&\cellcolor[HTML]{EFEFEF}1  &\cellcolor[HTML]{EFEFEF}73 (CTMC)     &\cellcolor[HTML]{EFEFEF}0.073 &\cellcolor[HTML]{EFEFEF}1.999953081e-6      &\cellcolor[HTML]{EFEFEF}&\cellcolor[HTML]{EFEFEF}37 (CTMC)  &\cellcolor[HTML]{EFEFEF}0.064&\cellcolor[HTML]{EFEFEF}2.9999929621e-6      \\

                  & \cellcolor[HTML]{EFEFEF}10  &    \cellcolor[HTML]{EFEFEF}73 (CTMC)    &  \cellcolor[HTML]{EFEFEF}0.076     &  \cellcolor[HTML]{EFEFEF}1.999530855e-5      & \cellcolor[HTML]{EFEFEF} & \cellcolor[HTML]{EFEFEF}37 (CTMC)   &\cellcolor[HTML]{EFEFEF}0.070   & \cellcolor[HTML]{EFEFEF}2.99929683e-5     \\

                  &100 &73 (CTMC)    &0.076      &0.0002001091927       &&37 (CTMC)      &0.071     &0.000300083976       \\
                    & \cellcolor[HTML]{EFEFEF}100000  &    \cellcolor[HTML]{EFEFEF}73 (CTMC)    &  \cellcolor[HTML]{EFEFEF}0.077     & \cellcolor[HTML]{EFEFEF}0.2772192934\textbf{*}     & \cellcolor[HTML]{EFEFEF} & \cellcolor[HTML]{EFEFEF}37 (CTMC)   & \cellcolor[HTML]{EFEFEF}0.074   & \cellcolor[HTML]{EFEFEF}0.3460009685\textbf{*}     \\

 \cmidrule(r){1-9} %\cmidrule(l){7-9} 
                      
\end{tabular}%
}
\begin{tablenotes}
\scriptsize
\item [*] * The reported probability for the reduced DFT is closer to the probability reported in \cite{Merle-thesis} for the same input failure distribution

\end{tablenotes}

\end{table}

\section{Conclusion }
\label{Conclusion}
In this work, we proposed a formal dynamic fault tree analysis methodology integrating theorem proving and model checking. We formalized the dynamic fault tree gates and operators in higher-order logic based on the time of failure of each gate. Using our formalization of the gates and the \texttt{extreal} library in HOL4, we proved over eighty simplification theorems that can be used to verify the reduction of any DFT. We used these theorems to verify the equivalence of the original and reduced DFT using theorem proving. In addition, we provided a formally verified qualitative analysis of the structure function in the form of reduced cut sets and sequences, which, to the best of our knowledge, is a novel contribution. The quantitative analysis of the reduced structure function is performed using model checking. This ensures that the model checking results correspond to the original DFT, since we use the formally verified reduced DFT in the quantitative analysis. Both the qualitative and the quantitative analyses were conducted on five benchmark DFTs, and the analysis results show that our proposed integrated methodology provides a formally verified reduced cut sets and sequences. In addition, the model checking results indicate that using the reduced DFT in the analysis has a positive impact on its cost in terms of both time and number of states. As a future work, we plan to provide the quantitative analysis of DFTs within the HOL theorem prover, which will allow us to have a complete framework for formal DFT analyses using theorem proving.

\balance
\newpage
\bibliography{mabiblio}{}
\bibliographystyle{IEEEtran}
\end{document}